# Gaming the Metrics? Bibliometric Anomalies and the Integrity Crisis in Global University Rankings

**Lokman I. Meho (0000-0003-0623-4652)***
**Department of Internal Medicine, American University of Beirut, Beirut, Lebanon**

**ABSTRACT:** This study explores how global university rankings incentivize institutions to prioritize bibliometric indicators at the expense of research integrity. Analyzing 98 fast-growing universities, we identified 18 with sharp declines in first and corresponding authorship roles, an early warning sign of potential metric manipulation. These institutions additionally exhibited unusually high publication surges in STEM fields, dense co-authorship networks, reciprocal citation behavior, and rising rates of self-citations, retractions, and publications in delisted journals. To assess these risks globally, the author developed the Research Integrity Risk Index ($RI^2$), a composite profiling metric combining retraction rates and reliance on delisted journals into a single score and ranking. The index effectively identifies institutions with atypical research performance patterns that deviate from global norms, highlighting systemic vulnerabilities in current evaluation frameworks. The findings underscore the need to rethink how research performance is measured to safeguard academic integrity and mitigate gaming behaviors.

## INTRODUCTION

Over the past two decades, global university rankings have profoundly reshaped academic priorities, institutional strategies, and the governance of higher education (Ahlers & Christmann-Budian, 2023; Hazelkorn & Mihut, 2021; Shen et al., 2023). Rankings from organizations such as the Academic Ranking of World Universities (Shanghai Ranking), QS, Times Higher Education, and U.S. News have elevated bibliometric indicators, such as publication and citation counts, to a central position in assessing institutional performance (Kochetkov, 2024; Rhein & Nanni, 2023; Sheeja et al., 2018). These metrics now influence everything from strategic planning and resource allocation to faculty hiring and funding decisions.

This transformation, however, has produced unintended consequences that can be understood through Goodhart's Law, which posits that "when a measure becomes a target, it ceases to be a good measure" (Dezhina, 2022; Fire & Guestrin, 2019). In academic research, once universities link publication and citation metrics to rankings or national prestige, they would no longer remain reliable indicators of research performance. Instead, these metrics can become susceptible to over-optimization, strategic inflation, and institutional behavior shifts prioritizing metric attainment over scholarly contribution (Hladchenko, 2025).

* To whom all correspondence should be made: lmeho@aub.edu.lb.

Evidence suggests that universities under reputational or performance pressures have engaged in metric-maximizing behaviors (Bhattacharjee, 2011; Biagioli et al., 2019; Catanzaro, 2024; Moosa, 2024; Pachter, 2014; Trung, 2020), including hyper-prolific authorship (Conroy, 2024; Ioannidis et al., 2024; Moreira et al., 2023), publication and citation clustering within closed institutional networks (Li et al., 2019; Silva et al., 2020), strategic publication in journals that offer visibility with minimal editorial gatekeeping (Gedik et al., 2024; Hladchenko, 2025; Maisonneuve, 2025; Teixeira da Silva, 2024), and other questionable authorship practices, such as gift or guest authorship (Morreim & Winer, 2023; Xie et al., 2021), ghost authorship (Pruschak & Hopp, 2022; Teixeira da Silva & Dobránszki, 2016), sold authorship (Abalkina, 2023; Chirico & Bramstedt, 2023; Ibrahim et al., 2025), and multi-affiliation inflation (Halevi et al., 2023; Hottenrott et al., 2021; Kuan et al., 2024). Although not all these practices are unethical, they cast doubt on the interpretability of bibliometric indicators and the extent to which rankings reflect genuine scholarly performance and impact. In extreme cases, these behaviors may erode the very value of the metrics the authors and institutions intended to maximize (Abalkina et al., 2025; Gruber, 2014; Oravec, 2017).

This study investigates a subset of universities whose bibliometric trajectories exhibit patterns consistent with the above concerns. Specifically, this study addresses the following research question: How can bibliometric anomalies be used to detect universities whose publication and citation patterns suggest metric-driven behaviors aimed at gaming global university rankings?

Through this analysis, the study highlights how performance pressures and metric-based incentives shape institutional behaviors in ways that challenge prevailing assumptions about academic excellence. While some of the observed trends may reflect legitimate capacity-building or internationalization efforts, others raise questions about the growing gap between what is counted and what counts (Olive et al., 2023). To address these concerns, the study seeks to inform policymakers, ranking agencies, universities, publishers, and authors of the significant risks posed by metric-driven distortions in academic evaluation. It underscores the urgent need for more transparent, accountable, and context-sensitive use of bibliometric indicators to ensure that research metrics serve as valid proxies for genuine scholarly contributions.

Finally, this study introduces a novel methodological framework that departs from conventional performance-based assessments by operationalizing bibliometric integrity risk through reproducible, resistance-to-gaming indicators. By focusing on structural vulnerabilities, rather than publication volume or citation counts, the study proposes a fundamentally different lens for evaluating institutional research behavior.

## MATERIALS AND METHODS

### Study Group Selection and Rationale

The study employed a multi-stage approach to identify universities with unusual trends in research output and behavior between the 2018-2019 and 2023-2024 calendar years—two periods selected to capture recent and rapid developments while minimizing transient disruptions from COVID-19 (Ioannidis et al., 2021; Raynaud et al., 2021; Zammarchi et al., 2024). Two-year intervals were used to provide a more stable basis for detecting sustained trends and to reduce the impact of short-term variability. The analysis focused on journal articles and reviews, excluding those with more than 100 co-authors to limit noise from mass collaborations.

Using SciVal data (May 1, 2025), the analysis began with the world's 1,000 most-publishing universities in 2023-2024, selected for their substantial research output and potential impact on global university rankings. From this group, 98 institutions, the extreme outliers, had publication growth exceeding 140% since 2018-2019, over five times the global average increase of 28%. Among these, 16 universities demonstrated sharp declines in research leadership roles, measured by first and corresponding authorship rates, which serve as recognized indicators of intellectual contribution and leadership (Chinchilla-Rodríguez et al., 2024; Kharasch et al., 2021). These universities showed declines exceeding 35% in first authorship and 15% in corresponding authorship over the study period, reflecting declines at least five times the global averages, where first authorship fell from 55% to 51%, and corresponding authorship from 54% to 52%. The remaining 82 universities exhibited an average decline of 2% (median = -1%) in first authorship rate and an 8% increase (median = +8%) in corresponding authorship rate. Diverging sharply from their peers, we selected the 16 institutions for in-depth analysis.

A parallel InCites analysis identified two additional institutions with bibliometric patterns consistent with the 16 universities from SciVal. This dual-source approach mitigated limitations associated with reliance on a single data source, strengthening the study's robustness. The final sample (study group) comprised 18 universities across India (7), Lebanon (1), Saudi Arabia (9), and the United Arab Emirates (1). University ages ranged from under 25 years (11 institutions) to over 50 years (3 institutions). Faculty sizes varied from fewer than 1,000 academics (5 institutions) to over 2,000 (7 institutions). Thirteen institutions included medical schools. Expanding the analysis to the top 1,500 universities would have identified 15 additional qualifying institutions (Table S1).

### Control Groups Selection and Rationale

The study employed national, regional, and international control groups to contextualize the study group's trends:

- **National Benchmark:** Each study group university was compared against its national aggregate research data (India, Lebanon, Saudi Arabia, and the United Arab Emirates) to assess whether observed trends reflected national shifts or institution-specific behaviors.

- **Regional Benchmark:** Four universities, one from each represented country, were selected based on consistent top-tier performance in global rankings (e.g., Shanghai Ranking): Indian Institute of Science (India), American University of Beirut (Lebanon), Khalifa University (United Arab Emirates), and King Abdullah University of Science and Technology (Saudi Arabia). These institutions served as stable, high-performing comparators.

- **International Benchmark:** Four top Shanghai Ranking institutions—Massachusetts Institute of Technology (MIT), Princeton University (Princeton), ETH Zurich, and the University of California, Berkeley (UC Berkeley)—were chosen for their global excellence and disciplinary similarity to the study group. The analysis excluded institutions with extreme disciplinary skew, such as Caltech (67% physics output) and Harvard, Oxford, and Stanford (over 50% medical publications). Study and control group universities maintained medical output near 15%, allowing fairer cross-institutional comparisons.

### Bibliometric Analysis

The study employed seven bibliometric indicators to determine whether the rapid increase in research output among the study group institutions resulted from organic academic development or from structural incentives and strategic publishing behaviors. Each indicator captures a distinct aspect of institutional publishing practices, with sources and analytical purposes specified below:

1. **Overall and Subject-Specific Research Output** (SciVal and Essential Science Indicators): The analysis measured institutional research productivity at both general and field-specific levels to determine whether growth patterns appeared broadly distributed or limited to specific disciplines.

2. **First and Corresponding Authorship Rates** (InCites): The study assessed shares of first and corresponding authorship to evaluate the extent of institutional involvement in research design, leadership, and communication, key indicators of intellectual contribution and accountability.

3. **Hyper-Prolific Authorship** (Elsevier, InCites, SciVal): The investigation examined extreme productivity patterns to detect potential authorship inflation or questionable publishing practices. Ioannidis et al. (2018) defined hyper-prolific authors (HPAs) as those with 72 or more publications in a calendar year, noting that 70% acknowledged not meeting the International Committee of Medical Journal Editors' (ICMJE) criteria in over a quarter of their publications. More recently, Moreira et al. (2023) characterized HPAs as "anomalous," citing unsustainable productivity patterns often linked to concentrated collaborative networks or recurring use of specific publication venues. This study adopts a threshold of 40 or more articles in a calendar year, balancing anomaly detection with disciplinary norms and legitimate variation in output across fields. HPAs were identified using three complementary methods:

- By Country: extracted from SciVal all authors in every country worldwide with 40 or more articles in any year from 2018 to 2024.
- By Field/Subfield: identified from SciVal the 500 most published authors in each year since 2018 across all 27 SciVal-defined fields and major subfields within broad domains such as engineering and medicine. Retained authors with 40 or more articles in any given year.
- By ORCID: downloaded from InCites a dataset of over 10,000 authors with ORCID IDs and 40 or more articles in any year from 2018 to 2024, matched them to SciVal records, and retained authors with 40 or more articles in SciVal.

  These approaches recovered 95% of Elsevier's independently compiled list (provided to the author upon request). The final dataset included 11,323 hyper-prolific authors, excluding articles with more than 100 co-authors. These authors represented 0.05% of all authors in SciVal during the study period, highlighting the rarity and possible ethical implications of such extreme publishing frequency.

4. **Articles in Delisted Journals** (Scopus and Web of Science): The author identified articles published in journals delisted by Scopus and Web of Science for ethical or quality-related violations to assess institutional reliance on questionable publishing venues (Wilches-Visbal et al., 2024).[1] Between 2009 and June 2025, a total of 978 unique journals were delisted—861 by Scopus and 169 by Web of Science. Of these, 555 were indexed in Scopus during 2018-2019 and accounted for 193,408 articles, and 209 were indexed in 2023-2024 with 125,465 articles. Institutional affiliations for these articles were tracked globally to evaluate exposure to low-integrity publication channels.

   **Retraction Rates** (Medline, Retraction Watch, Web of Science): The analysis used retraction data from three databases to evaluate institutional vulnerability to research misconduct or oversight failures. To do so, the author extracted and uploaded into SciVal all available original DOIs and PubMed IDs (PMIDs) associated with retracted articles from Retraction Watch, Medline, and Web of Science.

   As of June 18, 2025, Retraction Watch listed 43,482 entries marked as "Retraction" and classified under the following document types: case reports, clinical studies, guidelines, meta-analyses, research articles, retracted articles, review articles, letters when associated with research articles, and revisions when associated with review articles. These types were selected because, after cross-referencing, they corresponded to articles and reviews in Medline and Web of Science. Following the exclusion criteria used by Ioannidis et al. (2025), 2,238 records were removed due to non-author-related reasons (e.g., “Retract and Replace,” “Error by Journal/Publisher”).

[1] https://www.elsevier.com/products/scopus/content#4-titles-on-scopus and https://mjl.clarivate.com/collection-list-downloads

Of the remaining Retraction Watch entries, 38,316 successfully matched with SciVal records using DOIs and PMIDs. The remaining 5,166 could not be matched, either because they were published in journals not indexed by SciVal or lacked identifiable DOIs or PMIDs in the database.

To supplement the Retraction Watch dataset, an additional 4,416 unique retracted publications were identified from Medline and Web of Science (2,737 from Medline and 2,850 from Web of Science) that were classified as “Retracted” or “Retracted Publication” and tagged as articles or reviews. These did not overlap with any of the Retraction Watch entries.

In total, 42,732 unique retracted articles were matched to SciVal and included in the analysis. Scopus was excluded due to inconsistent classification practices: its “Retracted” label encompasses a broad range of document types, including letters and editorials, making it unsuitable for isolating retracted research articles and reviews.

To account for the time lag between publication and retraction, the analysis focused on articles published in 2022 and 2023, rather than 2023-2024, to better capture recent institutional behaviors while ensuring a broader view of retraction activity (Candal-Pedreira et al., 2024; Feng et al., 2024; Fang et al., 2012; Gedik et al., 2024). By June 18, 2025, the number of retracted articles stood at 10,579 for 2022, 2,897 for 2023, and 1,601 for 2024. Worldwide, as of June 18, 2025, the retraction rate for 2022-2023 averaged 2.2 per 1,000 articles, with the highest rates observed in mathematics (9.3) and computer science (7.6) and the lowest in arts and humanities (0.2) (see Table S2).

According to Retraction Watch, the 15 main reasons for retractions are: Investigation by Journal/Publisher (48%), Unreliable Results and/or Conclusions (42%), Investigation by Third Party (34%), Concerns/Issues About Data (30%), Concerns/Issues about Referencing/Attributions (26%), Paper Mill (25%), Concerns/Issues with Peer Review (23%), Concerns/Issues about Results and/or Conclusions (19%), Fake Peer Review (19%), Computer-Aided Content or Computer-Generated Content (18%), Duplication of/in Image (10%), Duplication of/in Article (8%), Euphemisms for Plagiarism (6%), Investigation by Company/Institution (6%), and Lack of IRB/IACUC Approval and/or Compliance (6%).

5. **Highly Cited Articles** (SciVal): To evaluate research visibility and reputational impact, the study relied on representation among articles ranked in the top 2% most cited globally. This threshold offers a broader and more discipline-inclusive measure than the top 1% benchmark commonly used in global rankings, such as the annual Clarivate's Highly Cited Researchers (Frietsch et al., 2025).

6. **Collaboration Patterns** (SciVal): Institutional partnerships were analyzed to determine whether growth in publication output coincided with changes in collaboration networks. Major collaborators were defined as external institutions that accounted for at least 2% of a university's total publications in a given period. This threshold captures sustained and potentially influential partnerships while filtering out incidental or short-term collaborations. Shifts in the number and identity of these collaborators from 2018-2019 to 2023-2024 were tracked and cross-referenced with Shanghai Ranking data to assess whether new or intensified partnerships aligned more with high- or lower-profile institutions.

## Data Sources

Bibliometric data were drawn primarily from:

- Scopus (Elsevier) and Web of Science (Clarivate): two of the world's largest multidisciplinary citation databases, indexing approximately 20 million and 18 million articles, respectively, between 2018 and 2024 (Baas et al., 2020; Birkle et al., 2020).

- SciVal and InCites: two analytics platforms built on Scopus and Web of Science databases, respectively, offering structured data on research output, citation performance, subject focus, authorship roles, and collaboration patterns.

# RESULTS

## Overall Research Output Surge

Between 2018-2019 and 2023-2024, universities in the study group showed substantial increases in research output, far exceeding national, regional, and international benchmarks. Institutions in India recorded publication growth from 243% to 766%, compared to a national average of 50%. The Lebanese American University (LAU) reported a 908% increase, while Lebanon's national growth rate increased by 102%, or 17% when excluding LAU. In Saudi Arabia, the national research output increased by 170%, or 120% when excluding study group universities, whereas the study group institutions grew between 220% and 965%. The University of Sharjah reported a 267% increase, surpassing the national average of 165% in the United Arab Emirates. In contrast, universities in the regional and international benchmark groups exhibited lower or stagnant growth, often falling below national trends (Table 1, Fig. S1).

**Table 1. Increases in research output of study and control group universities and their respective countries.** Institutions are represented by their acronyms or popular names. **India:** Chitkara=Chitkara University, CU=Chandigarh University, GLA=GLA University, IISc=Indian Institute of Science (Bangalore), KL=KL University, LPU=Lovely Professional University, SIMATS=Saveetha Institute of Medical and Technical Sciences, UPES=University of Petroleum and Energy Studies. **Lebanon:** AUB=American University of Beirut, LAU=Lebanese American University. **Saudi Arabia:** KAUST=King Abdullah University of Science and Technology, KKU=King Khalid University, KSU=King Saud University, NU=Najran University, PNU=Princess Nourah Bint Abdulrahman University, PSAU=Prince Sattam Bin Abdulaziz University, TU=Taif University, UOH=University of Ha'il, UQU=Umm Al-Qura University, UT=University of Tabuk. **Switzerland:** ETH Zurich. **United Arab Emirates:** KU=Khalifa University, UOS=University of Sharjah. **United States:** MIT=Massachusetts Institute of Technology, Princeton=Princeton University, UC Berkeley=University of California, Berkeley. Percentages for AUB and UC Berkeley are not shown due to respective declines of 5% and 2%. **Data Source:** SciVal (May 1, 2025). *The data for KL and NU in this Table are based on InCites.

| University (year founded) | Type | Number of articles published | | % of increase from 2018-2019 to 2023-2024 | | World ranking in # of articles | |
|---|---|---|---|---|---|---|---|
| **STUDY GROUP** | | 2018-2019 | 2023-2024 | Institution | Country | 2018-2019 | 2023-2024 |
| CU (India, 2012) | Private | 521 | 4,478 | 760% | 50% | 1500+ | 601 |
| Chitkara (India, 2010) | Private | 331 | 2,865 | 766% | 50% | 1500+ | 941 |
| GLA (India, 2010) | Private | 383 | 3,070 | 702% | 50% | 1500+ | 888 |
| KL (India, 1980)* | Private | 625 | 3,823 | 512% | 50% | 1500+ | 958* |
| LPU (India, 2005) | Private | 1,503 | 5,160 | 243% | 50% | 1164 | 514 |
| SIMATS (India, 2005) | Private | 3,037 | 10,418 | 243% | 50% | 656 | 187 |
| UPES (India, 2003) | Private | 489 | 2,662 | 444% | 50% | 1500+ | 1000 |
| LAU (Lebanon, 1924) | Private | 576 | 5,804 | 908% | 102% | 1500+ | 442 |
| KKU (Saudi Arabia, 1988) | Public | 2,070 | 11,096 | 436% | 170% | 905 | 166 |
| KSU (Saudi Arabia, 1957) | Public | 8,492 | 27,186 | 220% | 170% | 169 | 20 |
| NU (Saudi Arabia, 2006)* | Public | 380 | 2,450 | 545% | 170% | 1500+ | 990* |
| PSAU (Saudi Arabia, 2009) | Public | 1,245 | 8,519 | 584% | 170% | 1308 | 262 |
| PNU (Saudi Arabia, 1970) | Public | 818 | 8,709 | 965% | 170% | 1500+ | 247 |
| TU (Saudi Arabia, 2004) | Public | 996 | 5,166 | 419% | 170% | 1483 | 512 |
| UQU (Saudi Arabia, 1981) | Public | 1,124 | 5,704 | 407% | 170% | 1394 | 451 |
| UOH (Saudi Arabia, 2006) | Public | 534 | 3,118 | 484% | 170% | 1500+ | 875 |
| UT (Saudi Arabia, 2006) | Public | 705 | 3,350 | 375% | 170% | 1500+ | 823 |
| UOS (United Arab Emirates, 1997) | Public | 1,256 | 4,611 | 267% | 165% | 1303 | 576 |
| **REGIONAL GROUP** | | | | | | | |
| IISc (India, 1909) | Public | 4,124 | 4,556 | 10% | 50% | 480 | 587 |
| AUB (Lebanon, 1866) | Private | 2,234 | 2,113 | -5% | 102% | 855 | 1209 |
| KAUST (Saudi Arabia, 2009) | Private | 3,599 | 4,813 | 34% | 170% | 549 | 554 |
| KU (United Arab Emirates, 2017) | Public | 1,780 | 3,894 | 119% | 165% | 1029 | 700 |
| **INTERNATIONAL GROUP** | | | | | | | |
| ETH Zurich (Switzerland, 1855) | Public | 12,584 | 13,685 | 9% | 14% | 83 | 110 |
| MIT (United States, 1861) | Private | 15,115 | 15,528 | 3% | 1% | 56 | 83 |
| Princeton (United States, 1746) | Private | 6,896 | 7,443 | 8% | 1% | 245 | 325 |
| UC Berkeley (United States, 1868) | Public | 14,218 | 13,883 | -2% | 1% | 66 | 106 |

The research output gains among the study group universities led to a significant upward movement in global research rankings. In 2018-2019, only one institution in the study group ranked among the top 500 most-publishing universities. By 2023-2024, seven study group universities reached the top 500, and the median global rank for all 18 institutions improved from below 2,000th to 545th. Performance varied among regional comparators. For example, the Indian Institute of Science's research output grew by 10%, yet dropped from 480th to 587th in global publication rankings, while Khalifa University improved from just over 1,000th to 700th. Among international comparators, the median ranking fell from 75th to 108th, with two universities exiting the global top 100 (Table 1, Fig. S2).

In Shanghai Ranking, which places substantial weight on research output, the visibility of study group institutions also improved. In 2018-2019, only King Saud University from the group appeared in the top 1,000. By 2024, nine study group universities were listed (Fig. S3).

Several study group institutions made notable advances. Princess Nourah Bint Abdulrahman University debuted in the 301-400 range and the Lebanese American University first appeared in 2024, entering the 501-600 tier. Taif University and King Khalid University advanced by 600 and 400 positions, respectively, within two years of their initial inclusion in 2021 (Table S3). Only three other universities worldwide demonstrated similar rank mobility during this timeframe. In contrast, regional benchmark universities experienced stagnant or declining Shanghai Ranking performance while international benchmark universities retained their positions in the top global tiers, showing stable performance over time without fluctuations.

In India, where the National Institutional Ranking Framework (NIRF) is highly relevant to policy, none of the seven Indian study group institutions ranked among the top 40 before 2019, and only two were in the top 100. By 2024, four universities were ranked in the top 40 and six in the top 100.

### Inflation in Discipline-Specific Research

The increase in research output among study group universities becomes even more pronounced at the disciplinary level. In 2018-2019, only one study group university ranked among the world’s top 100 most-publishing institutions in any of the 21 subject categories tracked by Clarivate’s Essential Science Indicators. By 2023-2024, however, seven of the 18 study group universities collectively appeared 34 times among the top 100 across 13 subject categories. These gains occurred primarily in STEM fields, including chemistry, computer science, engineering, materials science, mathematics, pharmacology and toxicology, and physics (Fig. 1, Table S4).

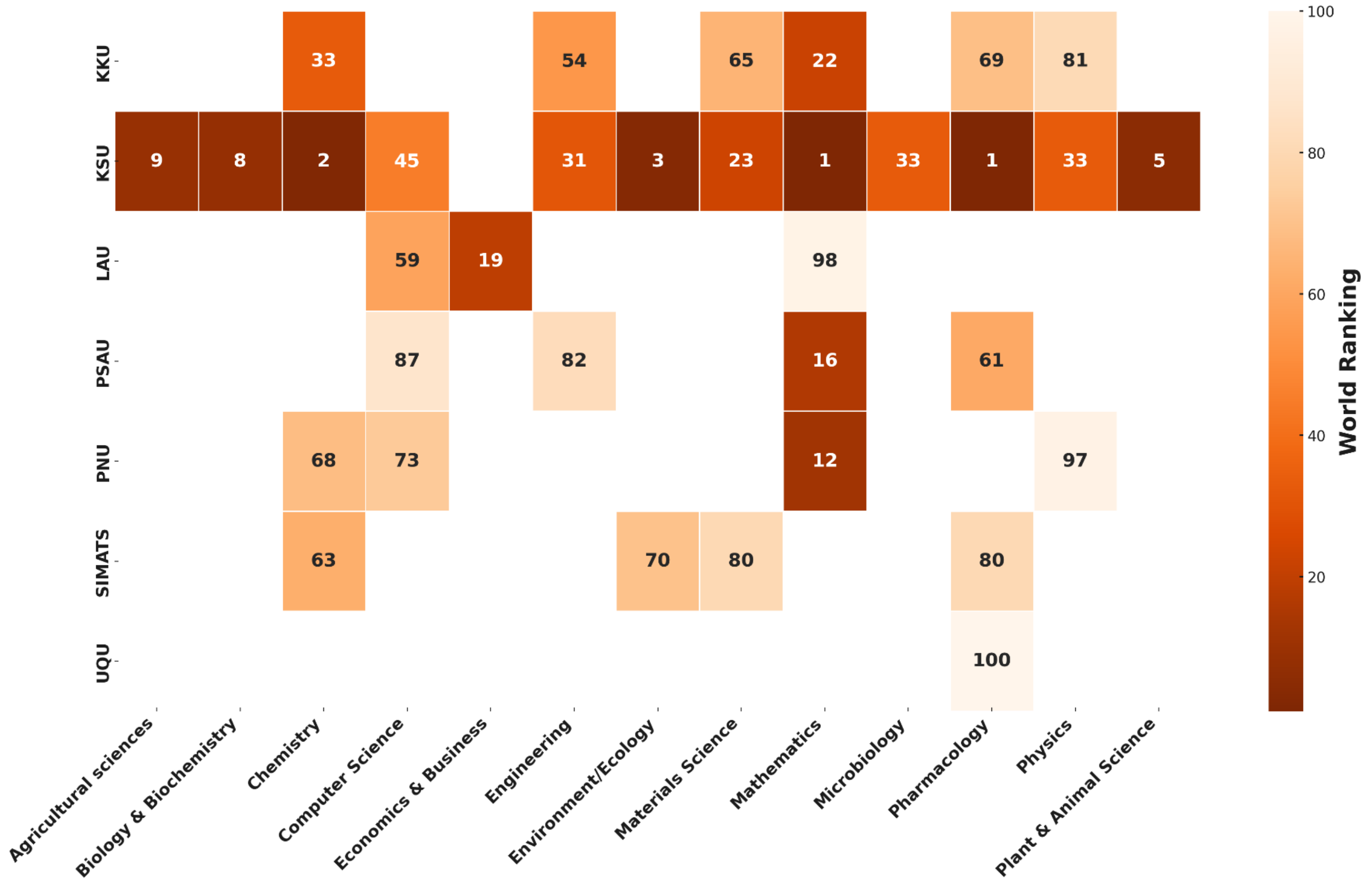

**Fig. 1. World Rankings of Study Group Universities Across Subject Categories (2023-2024, Top 100 Only).** KKU=King Khalid University, KSU=King Saud University, LAU=Lebanese American University, PNU=Princess Nourah Bint Abdulrahman University, PSAU=Prince Sattam Bin Abdulaziz University, SIMATS=Saveetha Institute of Medical and Technical Sciences, UQU=Umm Al-Qura University. **Data source:** Essential Science Indicators, via InCites (May 1, 2025).

A closer examination of these patterns highlights the extent of subject-specific concentration. For example, the Saveetha Institute of Medical and Technical Sciences, which had no representation among the world's top 1,000 most-published universities in any discipline in 2018-2019, appeared in four top-100 subject categories by 2023-2024. This figure surpassed all other Indian universities combined, except for the Vellore Institute of Technology, which also appeared in four categories. Similarly, the Lebanese American University, Prince Sattam Bin Abdulaziz University, and Princess Nourah Bint Abdulrahman University, none of which offer PhD programs, secured top-100 rankings on 11 occasions across seven subject categories.

Despite the substantial increases in overall research output, study group universities did not demonstrate similar growth in clinical medicine. This contrast is particularly notable given that 13 of the 18 study group institutions maintain sizable medical colleges, suggesting a growing disciplinary imbalance in research expansion (see Table S2).

Regional control group universities did not appear among the global top 100 in any subject category and showed minimal change in disciplinary representation during the study period. Meanwhile, international control institutions experienced a decline in disciplinary rankings, with collective appearances in the top 100 falling from 47 in 2018-2019 to 35 in 2023-2024, a reduction likely driven, in part, by the growing presence of study group universities in these top tiers.

### Decline in First and Corresponding Author Publications

First and corresponding authorship positions are widely recognized as indicators of intellectual leadership and responsibility in scientific research. These roles provide insight into where the core academic contributions originate within a publication. The analysis reveals a marked decline in the first and corresponding authorship rates among study group universities, diverging substantially from global and national patterns. Between 2018-2019 and 2023-2024, the average first authorship rate across these institutions fell by 44%, declining from 52% to 29%, more than six times the 7% global decline over the same period. Corresponding authorship rates decreased by 38%, from 52% to 32%, a decline over 13 times steeper than the 3% global average.

By 2023-2024, the decline was so steep that eight study group universities ranked among the ten lowest globally in first authorship rates, and 12 appeared in the bottom 20 (Table 2). The most extreme case was the Lebanese American University, where, based on InCites data, the first authorship rate dropped from 57% in 2018-2019 to 18% in 2023-2024, the sharpest decline recorded among the 1,000 most published institutions globally. A country-level breakdown further contextualizes these shifts. For example, among Indian institutions in the study group, first authorship rates declined by an average of 45% (median: 45%), compared to a national decline of 3.5%. Saudi institutions in the study group reported an average decrease of 46% (median: 40%), more than double the national decline of 21%.

Corresponding authorship rates followed similar trajectories, and authorship trends at the regional and international control institutions remained relatively stable, with only modest declines in first and corresponding authorship rates over the same period.

**Table 2. Decline in the proportion of, and world rank in, first and corresponding authorship rates. Data source:** InCites (May 1, 2025).

| | First authorship rate (%) | | | World ranking in first authorship rate | | Corresponding authorship rate (%) | | | World ranking in corresp. authorship rate | |
|---|---|---|---|---|---|---|---|---|---|---|
| **GROUP** | **2018-19** | **2023-24** | **Decline** | **2018-19** | **2023-24** | **2018-19** | **2023-24** | **Decline** | **2018-19** | **2023-24** |
| **STUDY** | | | | | | | | | | |
| CU | 59 | 31 | -48% | 343 | 993 | 52 | 35 | -31% | 610 | 981 |
| Chitkara | 66 | 32 | -51% | 211 | 988 | 56 | 40 | -28% | 452 | 915 |
| GLA | 72 | 38 | -47% | 93 | 960 | 68 | 38 | -44% | 160 | 962 |
| K L | 71 | 42 | -40% | 117 | 863 | 64 | 41 | -36% | 228 | 880 |
| LPU | 67 | 46 | -31% | 183 | 644 | 62 | 49 | -21% | 278 | 593 |
| SIMATS | 58 | 34 | -42% | 400 | 986 | 51 | 47 | -8% | 625 | 635 |
| UPES | 63 | 35 | -45% | 265 | 981 | 60 | 44 | -26% | 314 | 760 |
| LAU | 57 | 18 | -69% | 420 | 999 | 58 | 32 | -45% | 383 | 994 |
| KKU | 49 | 17 | -66% | 773 | 1000 | 48 | 16 | -66% | 787 | 1000 |
| KSU | 49 | 24 | -51% | 821 | 997 | 52 | 28 | -47% | 585 | 997 |
| NU | 44 | 45 | 3% | 957 | 712 | 53 | 33 | -39% | 542 | 993 |
| PSAU | 49 | 28 | -42% | 788 | 996 | 50 | 39 | -22% | 680 | 940 |
| PNU | 47 | 29 | -38% | 878 | 995 | 32 | 24 | -23% | 1000 | 998 |
| TU | 58 | 20 | -65% | 393 | 998 | 52 | 20 | -61% | 605 | 999 |
| UQU | 48 | 34 | -29% | 860 | 987 | 43 | 34 | -20% | 952 | 987 |
| UOH | 53 | 50 | -5% | 592 | 501 | 52 | 40 | -23% | 600 | 928 |
| UT | 52 | 31 | -41% | 621 | 992 | 49 | 29 | -40% | 760 | 996 |
| UOS | 51 | 39 | -24% | 702 | 939 | 57 | 48 | -16% | 410 | 625 |
| **REGIONAL** | | | | | | | | | | |
| IISc | 67 | 62 | -7% | 206 | 214 | 67 | 64 | -6% | 163 | 230 |
| AUB | 61 | 53 | -13% | 300 | 400 | 62 | 53 | -14% | 275 | 440 |
| KAUST | 50 | 47 | -7% | 610 | 625 | 57 | 54 | -5% | 405 | 430 |
| KU | 52 | 43 | -18% | 756 | 811 | 56 | 51 | -9% | 422 | 490 |
| **INTERNATIONAL** | | | | | | | | | | |
| ETH Zurich | 54 | 51 | -6% | 532 | 467 | 51 | 48 | -5% | 633 | 592 |
| MIT | 48 | 43 | -12% | 846 | 840 | 48 | 44 | -7% | 792 | 770 |
| Princeton | 52 | 47 | -10% | 618 | 600 | 50 | 48 | -5% | 680 | 630 |
| UC Berkeley | 50 | 44 | -13% | 740 | 762 | 48 | 44 | -9% | 767 | 793 |

### Rise in Hyper-Prolific Authorship

The number of HPAs, defined in this study as individuals publishing 40 or more articles in any calendar year, has risen globally. However, the increase accelerated most strikingly among the study group universities. Between 2018-2019 and 2023-2024, the number of HPAs affiliated with study group institutions rose by 1,252%, from 29 to 392: over 12 times the global increase (from 3,226 to 6,488 HPAs globally, excluding study group institutions), nearly eight times the 163% increase observed among regional control universities (from 8 to 21 HPAs), and fourteen times the 89% increase among the international control universities (from 9 to 17 HPAs) (Fig. S4).

The distribution of HPAs varied widely across institutions. King Saud University experienced the most significant increase, from four HPAs in 2018-2019 to 151 in 2023-2024, representing a 3,675% rise, the highest recorded globally (Fig. 2). An illustrative case is the Saveetha Institute of Medical and Technical Sciences (SIMATS), which highlights potential tensions between research quantity and quality. In 2018-2019, SIMATS had 16 HPAs, but 89% of their articles were published in journals later delisted from Scopus or Web of Science. Excluding those journals, none of the authors would have met the 40-article threshold used to classify HPAs.

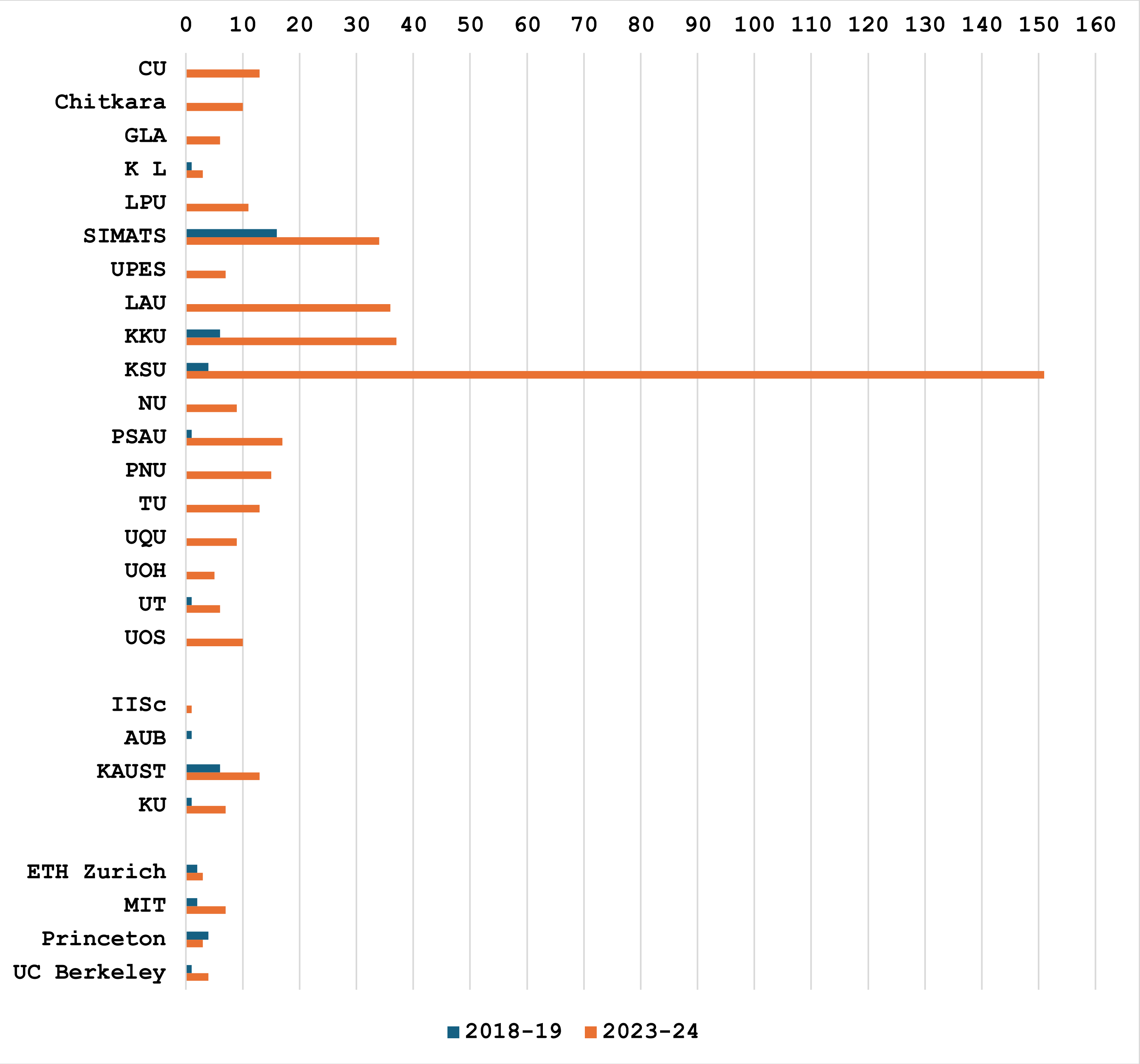

**Fig. 2. Changes in the number of hyper-prolific authors at universities in the study and control groups between 2018-2019 and 2023-2024. Data Source:** SciVal (May 1, 2025).

A defining characteristic of HPAs affiliated with study group institutions is their recent emergence. Among those identified in 2023-2024, 53% had never attained hyper-prolific status before, based on publication records dating back to 2004. This proportion is substantially higher than that observed in the regional comparison group (33%) and the international group (22%).

Significant increases in individual research productivity accompanied the sharp rise in hyper-prolific authorship at study group universities. The average annual output per HPA in these institutions rose from 12 articles in 2018-2019 to 53 in 2023-2024, a 342% increase. By contrast, HPAs in the regional control group increased by 68% (from 28 to 47 articles annually, and HPAs in the international group increased by 100% (from 26 to 52 articles). Table S5 illustrates three cases underscoring the diverse patterns of hyper-prolific authorship evident among study group universities.

**Publications in Delisted Journals and Retraction Patterns**

Between 2018-2019 and 2023-2024, 16 of the 18 study group universities reduced the share of their research output published in journals subsequently delisted by Scopus or Web of Science. For example, KL University and the Saveetha Institute of Medical and Technical Sciences decreased their proportions from over 75% to 15% and 8%, respectively. Among Saudi universities, the median share of output in delisted journals declined from 11% to 6% during the same period. By contrast, the Lebanese American University saw a nearly fourfold increase, and the University of Sharjah recorded a 33% rise in publications in delisted venues.

Despite the reduction, study group universities remain disproportionately represented among institutions with the highest publication volumes in delisted journals. In 2023-2024, they accounted for 15 of the top 50 universities globally by proportion of output in these venues. Numerically, King Saud University published 986 articles in delisted journals, the second-highest worldwide during this period (see Table S6).

In contrast, institutions in the regional and international control groups maintained consistently low engagement with delisted journals. None appeared in the global top 100 by either proportion or volume. On average, the share of regional control universities' articles in delisted journals comprised less than 1% of total output. International comparator institutions exhibited near-zero publication shares.

Beyond volume, the inclusion of articles from delisted journals in citation databases has implications for scholarly impact and research evaluation. While these journals have been removed from Scopus or Web of Science for not meeting editorial or publishing standards, their previously indexed articles continue to contribute to citation-based metrics. For example, although only 0.4% of articles in delisted journals are classified as highly cited, compared to 1.0% in indexed journals, they nonetheless account for 1% of all highly cited articles globally. This dynamic may influence outcomes in evaluation systems such as Clarivate's Highly Cited Researchers (HCR) list, which draws on cumulative citation data from the most recent 11 calendar years. Within this framework, even a small number of highly cited articles can substantially affect author- and institution-level visibility. In the present analysis, 284 authors and 87 institutions produced multiple highly cited papers in delisted journals. Given that the HCR list requires only 5-10 highly cited articles in several subject categories, the continued inclusion of delisted content may inadvertently affect eligibility and rankings.

A parallel concern is the steep rise in retraction counts and rates among study group institutions. As noted earlier, the 2022-2023 interval was selected as the reference period due to the delayed nature of retraction processes, which often unfold months or years after publication, making 2023-2024 data incomplete for comparison. Among the most striking cases, the Saveetha Institute of Medical and Technical Sciences recorded an increase in retractions from one in 2018-2019 to 154 in 2022-2023. King Saud University showed a similar trajectory, with retractions rising from 11 to 186 during the same period (see Fig. 3, Table S7).

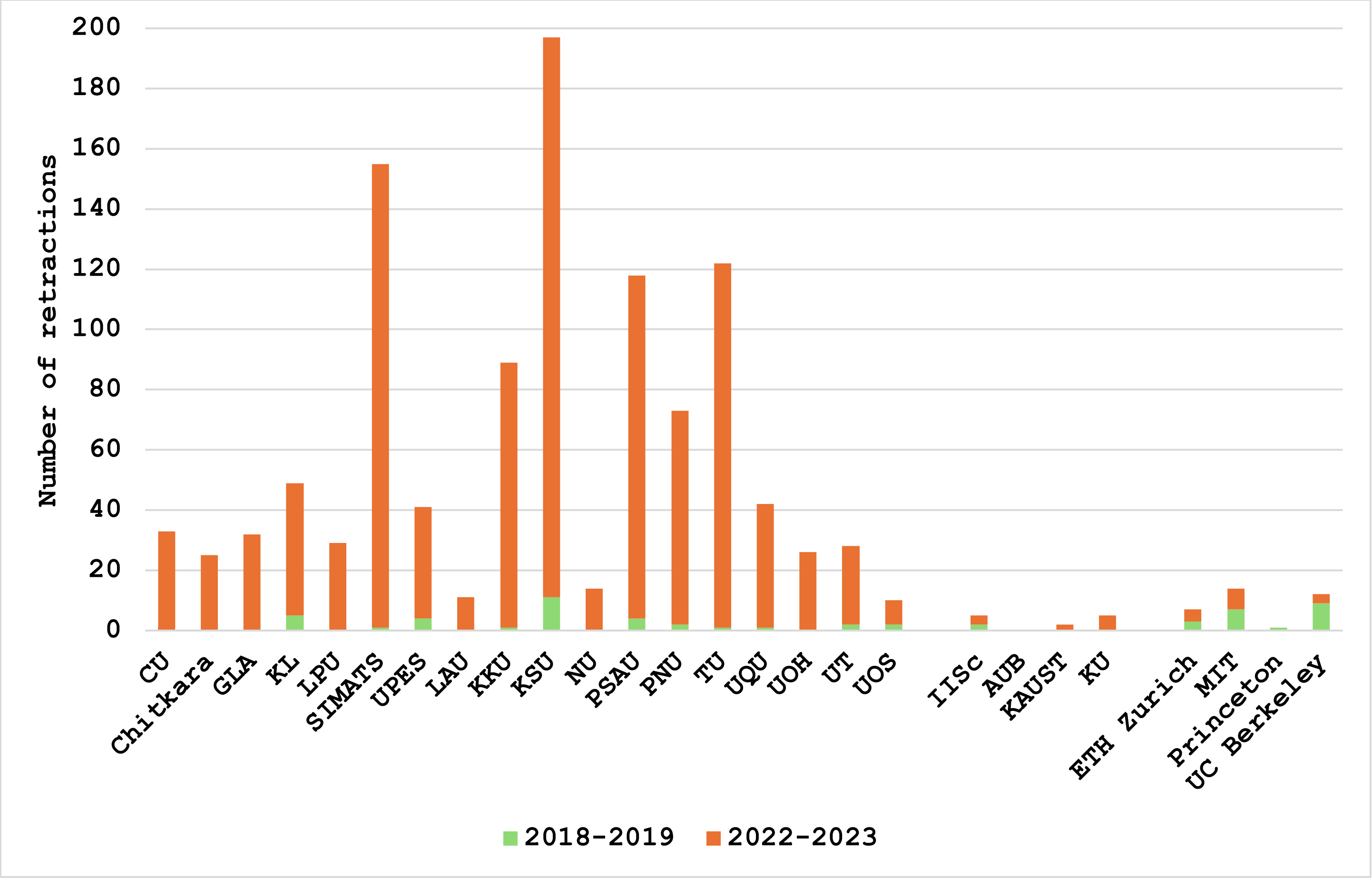

**Fig. 3. Retraction counts at the study, regional, and international group universities**

When normalized per 1,000 published articles, the study group's retraction rates were among the highest globally. In 2022-2023, they accounted for 15 of the top 50 highest-retraction-rate universities worldwide. This marks a dramatic shift from 2018-2019, when only one study group member ranked in the global top 50, and eight did not appear in the top 500. Regional and international control universities, on the other hand, maintained consistently low retraction rates across all analyzed years. Most recorded zero or near-zero retractions.

## Highly Cited Publications

Between 2018-2019 and 2023-2024, the number of study group universities' publications ranked among the world's top 2% most cited rose from 706 to 7,683 articles, representing a 988% increase. This growth outpaced the group's 326% increase in total research output, suggesting that factors beyond increased publication volume drove the rise in highly cited articles. By comparison, regional control universities recorded a 58% increase in highly cited articles against a 31% increase in total output. The international control group saw a more pronounced shift: while overall output rose by 4%, and highly cited publications fell by 25%.

The surge in highly cited articles was widespread among the study group. For instance, King Khalid University increased its highly cited output from 26 to 878 articles, a 3,277% rise, against a 436% increase in total production. KL University and the University of Petroleum and Energy Studies, which had no highly cited publications in 2018-2019, produced 205 and 265 such articles by 2023-2024, respectively (Table S8). Global standings in this category shifted substantially. In 2018-2019, none of the study group universities ranked among the world's top 200 institutions by share of articles classified as highly cited. By 2023-2024, 17 study group institutions entered the top 200 and nine ranked in the top 50. The most dramatic ascent was that of the Lebanese American University, which rose from 646th to 1st place globally.

This citation surge coincided with elevated self-citation rates among study group institutions, with a median of 6.0%, compared to 3.5-4.0% in the regional and international control groups (Table S9). The pattern also coincided with the emergence of dense inter-institutional citation networks within the study group. To examine these dynamics, we identified "major citation contributors" as institutions accounting for at least 1% of the citations received by a university's highly cited articles, a threshold determined through sensitivity analysis to balance relevance and noise. Cross-checking with overall citation data confirmed similar patterns, underscoring the robustness of the findings.

On average, each study group university had 17 major citation contributors in 2023–2024, including 9.5 from within the group, demonstrating substantial within-group cross-citation. As shown in Fig. 4, these interactions formed a dense web of bilateral or reciprocal citation activity. Beyond the group, 54 additional institutions were identified as significant contributors. Notably, none ranked in the global top 100; fewer than 10% appeared in the 101–500 range, and 67% were unranked by Shanghai Ranking. Frequently recurring external contributors included Al-Azhar University (Egypt), Graphic Era University (India), The Islamia University of Bahawalpur (Pakistan), Ural Federal University (Russia), Northern Border University (Saudi Arabia), and Yeungnam University (South Korea), each providing 1% or more of the citations to highly cited articles at five or more study group universities.

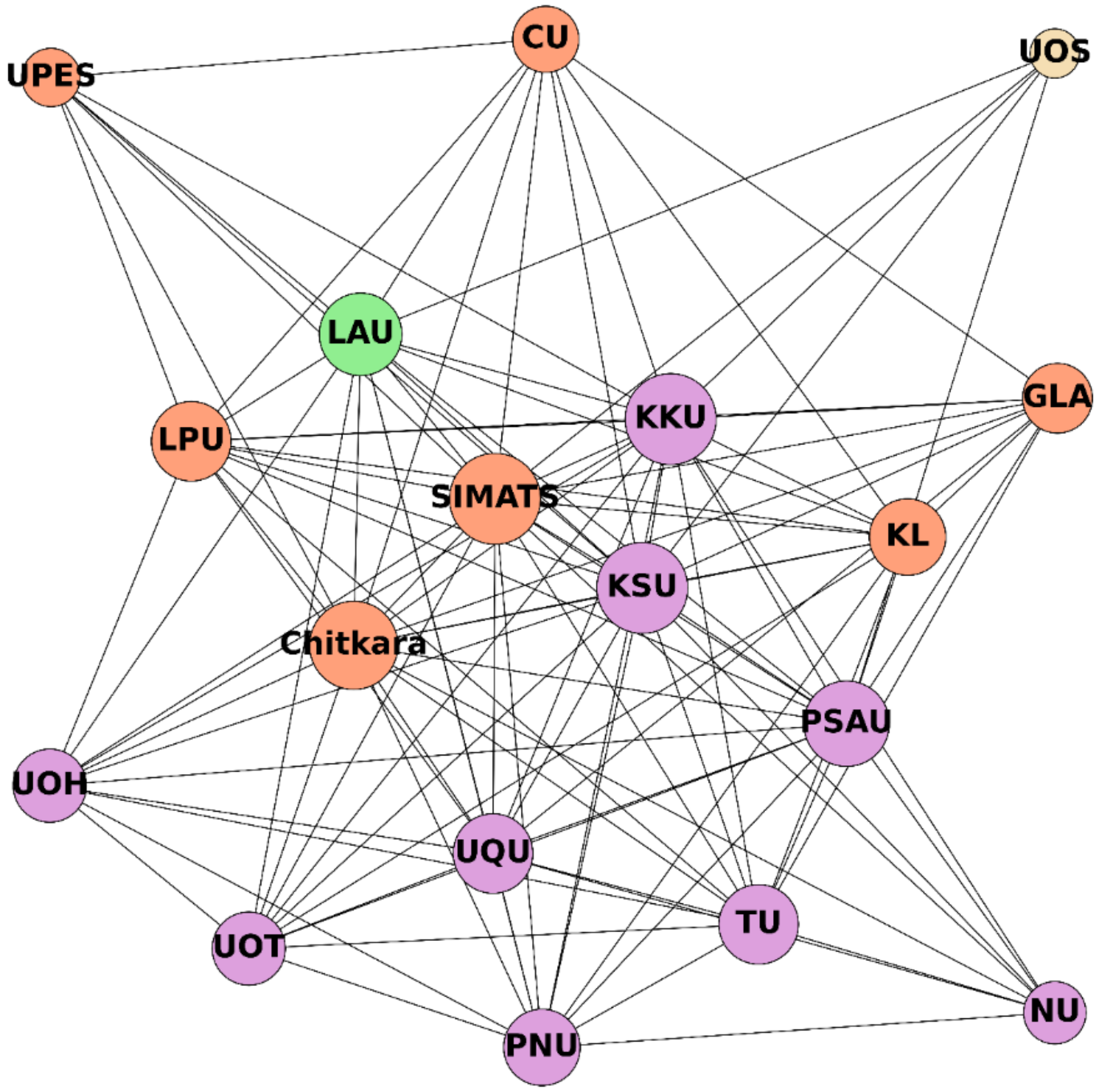

**Fig. 4. Internal network of major citation contributors to study group universities' highly cited publications, 2023-2024.** Node sizes represent the degree of each node, which refers to the number of direct connections (or links) a university has with others in the network. These connections are based on citation activity, where a link indicates that one university contributed to at least 1% of the citations received by another university's high-impact articles during the analyzed period, either unidirectionally or bidirectionally (reciprocally). Nodes with more connections appear larger, indicating greater centrality or influence within the citation structure. Node colors denote cluster membership, grouping institutions that are more densely interconnected. The layout is optimized to maximize spatial separation for visual clarity and highlight the underlying structure of inter-institutional citation relationships.

In contrast, regional and international control universities received citations from prominent institutions. Regional comparators averaged 14 major citation contributors, including Harvard, Tokyo, Toronto, and Tsinghua whereas international comparators had 22 major citation contributors, including Harvard, Cambridge, Stanford, Caltech, and Yale. Ninety-three percent of regional group contributors ranked among the global top 200, and 95% of international group contributors ranked in the top 100.

### The Role of External Contributors in Research Growth

To assess the extent to which external collaborations contributed to the study group's publication surge, we analyzed shifts in the number, identity, and academic standing of each university's primary co-authoring institutions, defined as contributors to at least 2% of a university's total publications in a given period.

Table S10 shows that the number of major collaborators expanded markedly. In 2018-2019, study group universities averaged ten major collaborators per institution. By 2023-2024, this number grew to 26. Of these, 15 were either new collaborators or previously minor contributors whose co-authorship share increased fivefold or more, indicating a substantial intensification of collaborative activity within a short timeframe.

Some institutions experienced especially pronounced shifts. For example, Saveetha Institute of Medical and Technical Sciences increased its major collaborators from one to 22, with 20 being new or significantly more engaged. The Lebanese American University demonstrated the most dramatic change, increasing from eight to 42 major collaborators, 39 of whom were new or had significantly intensified their collaboration levels between 2018-2019 and 2023-2024. By contrast, regional and international control universities exhibited only modest changes in their collaboration profiles.

Further analysis revealed that a substantial share of the new or intensified collaborators of study group universities were fellow study group members. On average, 41% of intensified collaborations occurred between peer institutions within the group. For instance, Chandigarh University reported 17 new partnerships, nine involving fellow study group universities (Table S11).

Fig. 5 visualizes this evolving internal network, comparing the study group's internal collaboration activity in 2018- 2019 with that of 2023-2024. The map shows limited to non-existent collaborations among the study group universities in 2018-2019. Indeed, six universities do not even appear on the map, indicating they had no notable partnership with each other or other study group members. Within five years, the network evolved into a dense web of connections representing significant bilateral collaborations. Many of the remaining collaborators outside this internal network were institutions with limited international academic visibility, totaling 95 in 2023-2024. Among these, 1% ranked among the global top 100 (based on the Shanghai Ranking), 5% were in the top 101-500, and 80% were unranked. Examples of frequently recurring partners included Future University in Egypt (Egypt), Anna University and Graphic Era (India), Jazan University and Northern Border University (Saudi Arabia), and Yeungnam University (South Korea), each serving as a major collaborator for five or more study group universities.

Notably, of the 425 authors whose publication history at the study group institutions began in 2023 and had contributed more than 10 publications each by May 2025, 159 (37%) held concurrent affiliations with two or more study group universities.

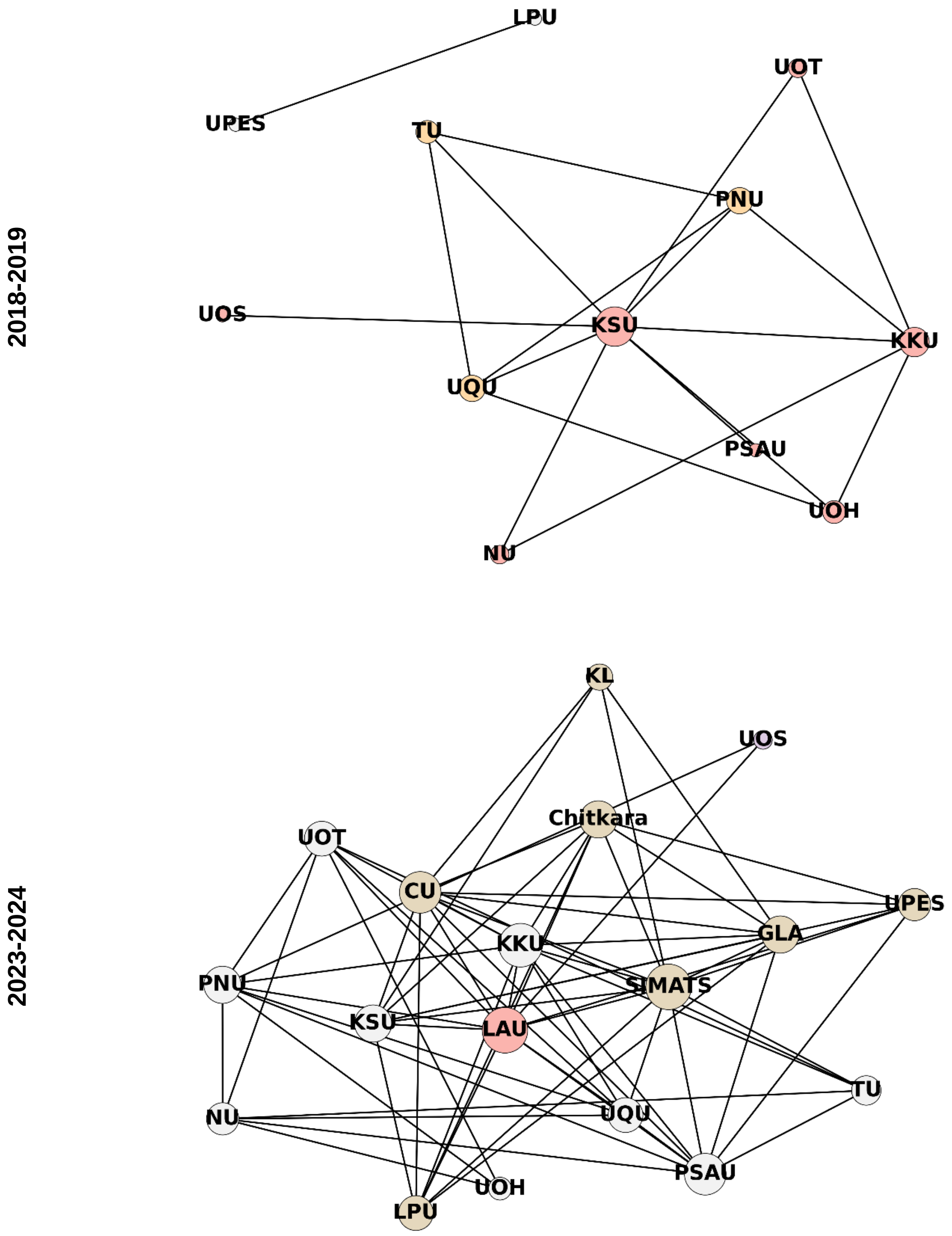

**Fig. 5. Collaboration network among study group universities (2018-2019 vs. 2023-2024).** Node sizes represent the degree of each node, which refers to the number of direct connections (or links) a university has with others in the network. These connections are based on co-authorship activity, where a link indicates that one university contributed to at least 2% of another university's total publications during the analyzed period, either unidirectionally or bidirectionally. Nodes with more connections appear larger, reflecting greater centrality or influence within the collaboration structure. Node colors indicate cluster membership, grouping institutions that are more densely interconnected. The layout is optimized to maximize spatial separation for visual clarity and highlight the underlying structure of inter-institutional relationships.

## The Research Integrity Risk Index (RI$^2$): A Composite Metric for Detecting Risk Profiles[2]

The convergence of multiple bibliometric anomalies among the study group institutions highlighted the need for a systematic framework to assess institutional-level research integrity risks. In response, this study introduces the Research Integrity Risk Index (RI$^2$), the first metric explicitly designed to profile such risks using empirically grounded, transparent indicators. Unlike conventional rankings that reward research volume and citation visibility, RI$^2$ shifts focus toward integrity-sensitive metrics that are resistant to manipulation and bibliometric inflation. In its current form, RI$^2$ comprises two primary components:

1. **Retraction Risk:** which measures the extent to which a university's research portfolio includes retracted articles, particularly those retracted due to data fabrication, plagiarism, ethical violations, authorship or peer review manipulation, or serious methodological errors (Fang et al., 2012; Ioannidis et al., 2025). In this study, it is calculated as the number of retractions per 1,000 articles over the most recent two full calendar years preceding the last one (e.g., 2022-2023 for an analysis conducted in 2025), normalizing for research volume and time-lag effects. Elevated retraction rates may signal structural vulnerabilities in institutional research governance, quality control mechanisms, and academic culture.

2. **Delisted Journal Risk:** which quantifies the proportion of an institution's publications that appear in journals removed from Scopus or Web of Science due to violations of publishing, editorial, or peer review standards (Cortegiani et al., 2020). This is measured over the most recent two full calendar years (e.g., 2023-2024 for an analysis conducted in 2025) and reflects structural vulnerabilities in quality control and publishing practices. Such publications continue to influence bibliometric metrics even after delisting, potentially distorting evaluative benchmarks.

Data for both indicators are drawn from Medline, Retraction Watch, and Web of Science, and serve as proxies for broader research integrity concerns, such as paper mills (businesses that sell authorship), citation cartels (reciprocal citation networks used to inflate impact), citation farms (organizations or networks that generate or sell citations), fraudulent authorship practices, and other forms of metric gaming (Abalkina, 2023; Maisonneuve, 2025; Candal-Pedreira et al., 2024; Feng et al., 2024; Ioannidis & Maniadis, 2024; Lancho Barrantes et al., 2023; Smagulov & Teixeira da Silva, 2025; Teixeira da Silva & Nazarovets, 2023; Wright, 2024). Importantly, both reflect verifiable outcomes rather than inferred behaviors, making them robust indicators of institutional-level risk.

To ensure consistency across diverse studies, whether analyzing 50 universities in one country or 5,000 globally, RI$^2$ applies a fixed reference group: the 1,000 most publishing universities worldwide. This global baseline provides balanced disciplinary and geographic coverage, ensuring thresholds are not skewed by outliers. It functions analogously to clinical reference ranges: just as hypertension is diagnosed using globally standardized thresholds, RI² classifications rely on a universal benchmark to detect structural anomalies. Key advantages of this approach include:

- **Universal comparability:** RI² scores are always interpreted against the global baseline, not rescaled to local or sample-specific norms. This ensures consistency across geographic and temporal contexts.

- **Stable thresholds:** Risk tiers are fixed based on the empirical distribution of the reference group. For example, if the highest observed retraction rate is 3 per 1,000 articles, an institution with 1.5 retractions receives a normalized score of 0.5, regardless of sample size.

[2] https://sites.aub.edu.lb/lmeho/ri2/.

### Normalization and Composite Scoring

Each indicator is scaled to a 0–1 range using Min-Max normalization relative to the global reference group. The composite RI² score is the simple average of the two:

*RI2 = (Normalized Retraction Rate + Normalized Delisted Rate) / 2*

For the June 2025 edition, for example, the retraction rates ranged from 0.00 to 26.82 retractions per 1,000 articles, and the share of articles in delisted journals ranged from 0.00% to 15.35%. These values define the normalization scale and remain fixed across all samples, ensuring stable cross-institutional comparisons regardless of geographic or disciplinary representation.

### Fixed Tier Classifications

Institutions are then classified into one of five risk tiers based on their RI² score (Fig. 6).

| Risk Tiers | Percentile range | Interpretation | Score range (June 2025 edition) |
|---|---|---|---|
| **Red Flag** | **≥ 95th** | **Extreme anomalies; systemic integrity risks** | **RI² ≥ 0.252** |
| **High Risk** | 90 - 95th | Significant deviation from global norms | 0.174 ≤ RI² < 0.252 |
| **Watch List** | 75 - 90th | Moderately elevated risk; emerging concerns | 0.099 ≤ RI² < 0.174 |
| **Normal Variation** | 50 - 75th | Within expected global variance | 0.049 ≤ RI² < 0.099 |
| **Low Risk** | < 50th | Strong adherence to publishing integrity norms | RI² < 0.049 |

**Fig. 6. RI² Risk Tiers Framework (based on the 1,000 most publishing universities, 2023-2024)**

### Key Features of the RI² Methodology

In summary, key features of the RI² methodology include:

- **Global benchmarking:** Ensures representative and statistically reliable classification.

- **Fixed thresholds for consistency:** Applicable across large or small datasets, preserving interpretability.

- **Transparency and robustness:** Built on normalized, verifiable metrics rather than subjective assessments.

When applied to the 1,000 most publishing universities examined in this study, RI² proved effective in distinguishing risk profiles. Seventeen of the 18 study group universities fell into the Red Flag tier, and one into High Risk. In contrast, six of the eight control institutions were classified as Low Risk, with two under Normal Variation. These findings support RI²'s utility as a scalable, evidence-based tool for identifying institutional research integrity risks that traditional productivity metrics may overlook.

## DISCUSSION

The findings in this study raise important questions about the extent to which global university rankings, through their emphasis on bibliometric indicators, may inadvertently incentivize behavior that undermines research integrity. While the observed publication surges and citation gains among the study group universities might signal growing research capacity or internationalization, the accompanying decline in first and corresponding authorship roles, increased reliance on delisted journals, dense reciprocal collaboration and citation networks, and rising retraction rates suggest that more complex and potentially problematic dynamics are at play.

The magnitude and speed of publication growth among the study group universities, often exceeding 400% within five years, raise questions about whether such expansion can be attributed to organic academic development alone. While institutional investments may partially explain this growth as is the case with China's recently founded Southern University of Science and Technology (2011) and Westlake University (2018), the anomalous bibliometric patterns, concentration in STEM fields, and absence of similar gains in disciplines such as clinical medicine and the social sciences among the study group suggest more strategic, ranking-oriented expansion. This abrupt rise indicates a deliberate focus on maximizing bibliometric output.

The decline in first and corresponding authorship, key indicators of intellectual leadership, further supports the possibility of strategic behavior. In contrast to global baselines, study group universities showed sharp and consistent declines in these roles. While such trends might stem from international collaborations, their scale and uniformity suggest deeper structural changes. Cases documented by Abalkina (2023) and Vasconez-Gonzalez et al. (2024) provide illustrative parallels, including the influence of authorship-for-sale schemes, AI-generated content, and paper mills. These patterns raise concerns about institutions prioritizing metrics over meaningful scholarly contribution (Maisonneuve, 2025).

The proliferation of HPAs, often newly affiliated with study group universities (Meho & Akl, 2025), reinforces these concerns. Many HPAs listed multiple concurrent affiliations, thereby amplifying perceived international collaboration and inflating output statistics. While not inherently unethical, such affiliations may misrepresent actual academic engagement. Prior studies show how multi-affiliations can distort bibliometric indicators and facilitate symbolic co-authorship (Gök & Karaulova, 2024; Hottenrott et al., 2021; Ioannidis & Maniadis, 2024; Lin et al., 2025). Without transparency on contribution roles, these practices complicate the interpretation of collaboration metrics.

Concerns about quality are further underscored by continued reliance on delisted journals. Although some institutions improved their publishing practices over time, many remained among the top contributors to journals excluded from major databases for poor editorial standards (Cortegiani et al., 2020). These same universities also saw a steep rise in retractions, a trend not observed among comparators. As noted by Ioannidis et al. (Ioannidis et al., 2025), retracted authors often display high self-citation, limited career duration, and inflated productivity, profiles that closely mirror those of several HPAs in this study.

Citation and collaboration patterns reveal additional anomalies. Study group universities had higher self-citation rates and frequently cited each other, forming dense internal citation networks. The rise in highly cited articles suggests potential citation coordination. These behaviors echo broader trends of citation cartels, coercive citations, and citation-for-sale practices that distort impact metrics (Baccini & Petrovich, 2023; Chawla, 2024; Ibrahim et al., 2025; Ioannidis & Maniadis, 2024; Kojaku et al., 2021). Major collaborators also surged, often lacking global academic visibility. By 2023-2024, 80% of frequent partners were unranked, raising questions about the strategic nature of these alliances. Notably, co-affiliation patterns among authors, particularly those linking multiple study group institutions, have created ecosystems optimized for mutual metric enhancement.

Historical examples highlight the risks of metric-centric strategies in academic evaluation. In Vietnam, Duy Tan and Ton Duc Thang Universities experienced extraordinary publication growth between 2016 and 2020, securing positions within the 601-700 band of the Shanghai Ranking by 2021. However, subsequent investigations revealed that some of this surge was driven by controversial authorship practices, including the strategic extension of secondary affiliations to foreign researchers (Trung, 2020). Following public scrutiny and regulatory intervention (Oransky, 2024), both institutions saw steep declines in output and were excluded from the Shanghai Ranking by 2023.

Parallel trajectories have begun to emerge within the study group. Taif University, for instance, expanded its publication output from 480 articles in 2018 to 4,673 in 2022, before dropping to 2,356 in 2023 and recovering slightly to 2,807 in 2024. Its Shanghai Ranking placement fell from the 201-300 band in 2023 to the 401-500 range in 2024. The Lebanese American University demonstrated a comparable pattern, producing only 27% of its 2024 total output in the first five months of 2025, well below the national average of 36%, the regional average of 45%, and the global average of 40%. These examples illustrate a familiar arc: meteoric rise driven by aggressive metric accumulation, followed by rapid decline when scrutiny, policy changes, or internal limits are reached. Institutions that prioritize short-term bibliometric gains without parallel investments in research infrastructure, governance, and quality assurance risk reputational damage, output stagnation, and long-term academic instability.

Global university rankings continue to shape institutional behavior by emphasizing volume- and citation-based metrics, often at the expense of validity and ethical robustness. As Teixeira da Silva (2024) observes, many ranking systems lack safeguards and rarely impose meaningful sanctions, even when manipulation is exposed. A notable case involved a voting syndicate among 21 Arab universities, which was uncovered by Times Higher Education (Watkins, 2024). Despite confirming coordinated behavior, THE merely neutralized the affected votes and retained all institutions in the rankings, highlighting the system's tolerance for superficial compliance over substantive reform.

Media investigations have further revealed how current ranking methodologies can reward citation inflation and questionable research practices. For example, Indian universities flagged in this study were ranked above the Indian Institute of Science, prompting public backlash and scrutiny of THE's reliance on self-citation-inflated metrics like the Field-Weighted Citation Impact (Agrawal, 2024). These cases underscore the urgent need to realign ranking methodologies to reflect integrity as well as output.

Tools like the $RI^2$ offer empirically grounded mechanisms to identify high-risk institutions and flag potential metric manipulation. Embedding such tools into global ranking systems can enhance the reliability and legitimacy of academic evaluation by highlighting integrity-sensitive outliers, rather than merely rewarding publication scale or citation volume.

Unlike other tools, $RI^2$ introduces a scalable, rank-based model grounded in quantitative institutional-level data. This methodological pivot represents a conceptual advancement by enabling early detection of integrity risks in ways that are transparent, conservative, and scalable across diverse academic systems. $RI^2$ introduces a shift from performance-based to risk-based evaluation and profiling, relying on hard-to-game research indicators. Its deliberately conservative architecture minimizes inflationary bias and emphasizes specificity, offering a more credible lens through which to assess institutional behavior.

The index not only validates previously flagged anomalies but also reveals additional at-risk institutions overlooked by conventional metrics. As a red-flag mechanism, RI$^2$ makes structural vulnerabilities visible, often masked by inflated bibliometric profiles.

Designed as a dynamic tool, RI$^2$ may in the future incorporate additional dimensions while adjusting for institutional size, age, region, and disciplinary profile. In doing so, it provides a scalable and interpretable benchmark that complements existing evaluation frameworks.

Ultimately, RI$^2$ promotes accountability by discouraging overreliance on low-integrity publishing venues, exposing sophisticated manipulation tactics, and pressuring bibliometric databases to uphold stricter inclusion standards. Its adoption would mark a significant advance toward restoring trust in global academic rankings.

Finally, this study have implications for many stakeholders. Universities must rebalance incentives by rejecting metric-driven strategies and aligning with COPE and ICMJE guidelines (Moher et al., 2020). Ethical authorship criteria, transparency in affiliations, and internal audits of publication practices are essential. Sustainable growth requires investment in research capacity and responsible mentorship.

Faculty are vital to upholding research integrity. Declining lead authorship and rising symbolic affiliations signal weakening institutional ownership of scholarship. Faculty must reinforce norms by mentoring juniors, resisting metric pressure, and supporting transparent authorship. They can lead governance reforms, such as authorship audits and clearer multi-affiliation policies, especially in fast-growing or externally funded environments.

Governments and funders should reconsider how metrics shape research agendas. Programs like Saudi Arabia’s Vision 2030 and India’s NIRF promote competition but risk overemphasizing volume. Institutions with recurring anomalies should face review and funding consequences. Conversely, integrity-centered practices should be rewarded. Cross-border funding tied to Open Access publishing fees and article processing charges and multi-affiliations must be auditable to ensure alignment with scholarly contributions.

Bibliometric platforms must improve quality assurance, including ORCID-based verification and alerts for abnormal behavior. Journal retention criteria must be transparent. Despite delisting some journals, others with known quality issues remain indexed, inflating metrics. Publishers must detach business incentives from editorial oversight. Accreditation agencies should embed integrity audits and consider conditional accreditation for institutions with persistent red flags.

Inflated outputs from a few institutions, often bolstered by highly cited articles in delisted journals, are reshaping global benchmarks, disadvantaging integrity-focused scholars. From 2018-2019 to 2023-2024, HPAs among study group universities rose from 29 to 392, and top 2% cited articles jumped from 706 to 7,683. Such distortion raises thresholds for recognition (e.g., Highly Cited Researchers), marginalizing rigorous, reproducible research and distorting merit-based assessments.

Students and families often equate rankings with quality, yet rankings that reward volume send unreliable signals. Institutions may prioritize visibility over intellectual substance. As global benchmarks shift, deserving institutions may struggle for recognition. For example, inflating citations can affect eligibility for groups like the Association of American Universities, which uses citation-based metrics and funding as key criteria.

## LIMITATIONS AND DIRECTIONS FOR FUTURE RESEARCH

This study has several limitations. First, it is diagnostic, not causal; however, its purpose is to identify reproducible patterns of bibliometric risk, not to establish causation. Second, the study group was selected using high-sensitivity thresholds for research growth and declines in lead authorship roles. The broader application of RI² to the 1,000 most-publishing universities confirms the wider relevance of the observed risk patterns. Third, the analysis relies on bibliometric indicators from Scopus, Web of Science, and Retraction Watch. These indicators, while empirically linked to integrity concerns, cannot capture full ethical nuance. Anomalies should be seen as signals of systemic misalignment, not proof of misconduct. Fourth, the 40-article threshold for HPAs, although grounded in prior literature and suited for STEM, may underrepresent norms in other fields. Future adjustments should consider disciplinary contexts. Fifth, delisting practices are inconsistent and opaque. The study used the union of delisted journals from Scopus and Web of Science to enhance coverage, acknowledging gaps in indexing oversight. Sixth, retraction rates are lagging indicators. To balance timeliness and accuracy, the analysis used a two-year lag for recent data. Seventh, RI² currently captures only three risk dimensions, using a rank-sum approach. While this design ensures simplicity and robustness, it omits other vulnerabilities (e.g., self-citation, multi-affiliations) and disciplinary norms, which future iterations should address.

Future research should expand and validate the RI² framework, incorporating additional indicators. Enhancements like weighted scoring, field normalization, and discipline-specific thresholds will improve fairness and benchmarking precision. Longitudinal studies are needed to determine whether recent publication surges are sustainable or fragile. Qualitative research with scholars and administrators could shed light on the drivers behind metric-driven behavior. Comparative policy studies should explore how evaluation models shape institutional strategies, helping align research incentives with integrity, mentorship, and societal value.

**Acknowledgments:** The author gratefully acknowledges Debora Shaw for her feedback on the manuscript. The author also thanks Elie A. Akl for his co-authorship on prior related research, which provided a valuable starting point for this study.

**Funding:** None

**Author contributions:**

Conceptualization, Methodology, Investigation, Visualization, Supervision, Writing (original draft, review, and editing)

**Competing interests:** The author declares an institutional affiliation with an organization located in the same country as one of the universities examined in the study. This affiliation has not influenced the design, analysis, or interpretation of the findings presented.

**Data and materials availability:** Data can be accessed at https://sites.aub.edu.lb/lmeho/ri2/.

# Supplementary Materials for

## Gaming the Metrics? Bibliometric Anomalies and the Integrity Crisis in Global University Rankings

## Table S1. Additional Universities Meeting Selection Criteria (Top 1001-1500 Analysis)

| Name | Country | Articled in delisted journals, 2018-2019 | Articled in delisted journals, 2023-2024 | Retractions rate per 1,000 articles, 2018-2019 | Retractions rate per 1,000 articles, 2022 | Retraction rate per 1,000 articles, 2023-2024 | Top 2% most cited articles, 2018-2019 (%) | Top 2% most cited articles, 2023-2024 (%) | Articles, 2018-2019 | Articles, 2023-2024 | Research growth from 2018-2019 to 2023-2024 (%) | First authorship rate, 2023-2024 (%) | Change in 1st authorship rate from 2018-19 to 2023-24 | Corresponding authorship rate, 2023-2024 (%) | Change in corresponding authorship rate from 2018-19 to 2023-24 | Hyper-prolific authors, 2023-2024 |
|---|---|---|---|---|---|---|---|---|---|---|---|---|---|---|---|---|
| **DIU** | Bangladesh | 121 (46%) | 149 (9%) | 8 | 30 | 5 | 2.3% | 7.6% | 263 | 1,649 | 527% | 24 | -47% | 29 | -12% | 5 |
| **FUE** | Egypt | 11 (5%) | 78 (4%) | 9 | 21 | 8 | 0.4% | 9.1% | 235 | 1,953 | 731% | 13 | -73% | 25 | -32% | 6 |
| **GEU** | India | 28 (16%) | 169 (7%) | 0 | 80 | 5 | 4.7% | 8.4% | 171 | 2,521 | 1374% | 24 | -58% | 41 | -22% | 6 |
| **UU** | India | 41 (36%) | 94 (6%) | 9 | 129 | 5 | 0.9% | 9.5% | 113 | 1,707 | 1411% | 13 | -80% | 17 | -74% | 6 |
| **MUS** | Iraq | 77 (59%) | 157 (6%) | 0 | 40 | 15 | 0.8% | 7.2% | 130 | 2,428 | 1768% | 11 | -72% | 16 | -59% | 7 |
| **AAU** | Jordan | 51 (27%) | 168 (10%) | 0 | 2 | 2 | 0.5% | 7.8% | 188 | 1,691 | 799% | 24 | -67% | 33 | -53% | 3 |
| **ASU** | Jordan | 40 (17%) | 246 (10%) | 0 | 5 | 6 | 0.4% | 6.2% | 243 | 2,378 | 879% | 27 | -53% | 17 | -60% | 5 |
| **MEU** | Jordan | 31 (29%) | 259 (14%) | 0 | 0 | 5 | 0.9% | 8.3% | 108 | 1,795 | 1562% | 26 | -48% | 27 | -43% | 4 |
| **INTI** | Malaysia | 46 (35%) | 179 (11%) | 0 | 0 | 1 | 0.8% | 3.8% | 133 | 1,624 | 1121% | 25 | -44% | 35 | -6% | 8 |
| **SU** | Malaysia | 39 (5%) | 120 (5%) | 0 | 1 | 2 | 5.4% | 11.9% | 735 | 2,426 | 230% | 22 | -38% | 36 | -27% | 10 |
| **UMT** | Pakistan | 23 (4%) | 99 (5%) | 2 | 25 | 2 | 2.7% | 9.8% | 554 | 1,807 | 226% | 47 | -4% | 39 | -17% | 2 |
| **NBU** | Saudi Arabia | 45 (13%) | 209 (10%) | 3 | 18 | 3 | 0.3% | 5.0% | 356 | 1,993 | 460% | 36 | -52% | 39 | -41% | 2 |
| **YZU** | Taiwan | 24 (3%) | 50 (3%) | 1 | 0 | 2 | 1.1% | 8.5% | 731 | 1,820 | 149% | 21 | -59% | 39 | -32% | 7 |
| **IU** | Turkey | 9 (3%) | 53 (3%) | 0 | 7 | 2 | 0.4% | 5.5% | 269 | 1,907 | 609% | 23 | -35% | 35 | 4% | 6 |
| **AU** | United Arab Emirates | 43 (20%) | 150 (7%) | 0 | 28 | 5 | 0.5% | 5.5% | 216 | 2,223 | 929% | 22 | -50% | 36 | -25% | 11 |

Acronyms listed in the order they appear in the Table: DIU=Daffodil International University, FUE=Future University in Egypt, GEU=Graphic Era University, UU=Uttaranchal University, MUS=Al-Mustaqbal University College, AAU=Al Ahliyya Amman University, ASU=Applied Science Private University, MEU=Middle East University, Jordan, INTI=INTI International University, SU=Sunway University, UMT=University of Management and Technology, NBU=Northern Borders University, YZU=Yuan Ze University, IU=Istinye University, AU=Ajman University. **Data Source:** SciVal (May 1, 2025). All except Saudi Arabia's NBU are private universities. None had HPAs in 2018-2019 except Malaysia's SU, which had one.

**Table S2. Retraction rates by subject category, 2022-2023**

| Subject Area | Retracted articles | Total articles | Retraction rate (per 1,000 articles) |
|---|---|---|---|
| Mathematics | 3,558 | 384,131 | 9.3 |
| Computer Science | 4,408 | 582,208 | 7.6 |
| Immunology and Microbiology | 1,626 | 220,234 | 7.4 |
| Neuroscience | 964 | 164,502 | 5.9 |
| Biochemistry, Genetics and Molecular Biology | 3,330 | 752,898 | 4.4 |
| Engineering | 4,336 | 1,097,322 | 4.0 |
| Multidisciplinary | 342 | 143,580 | 2.4 |
| Decision Sciences | 130 | 704,58 | 1.8 |
| Medicine | 3,102 | 1,705,380 | 1.8 |
| Environmental Science | 889 | 552,401 | 1.6 |
| Chemical Engineering | 609 | 419,827 | 1.5 |
| Physics and Astronomy | 877 | 645,154 | 1.4 |
| Energy | 282 | 283,905 | 1.0 |
| Chemistry | 622 | 636,043 | 1.0 |
| Health Professions | 95 | 112,008 | 0.8 |
| Materials Science | 579 | 691,768 | 0.8 |
| Business, Management and Accounting | 130 | 168,869 | 0.8 |
| Economics, Econometrics and Finance | 92 | 125,613 | 0.7 |
| Pharmacology, Toxicology and Pharmaceutics | 159 | 225,582 | 0.7 |
| Psychology | 110 | 168,806 | 0.7 |
| Social Sciences | 424 | 684,227 | 0.6 |
| Agricultural and Biological Sciences | 212 | 560,047 | 0.4 |
| Nursing | 44 | 129,299 | 0.3 |
| Earth and Planetary Sciences | 90 | 295,721 | 0.3 |
| Dentistry | 12 | 40,236 | 0.3 |
| Veterinary | 14 | 60,974 | 0.2 |
| Arts and Humanities | 55 | 297,328 | 0.2 |
| **TOTAL** | **13,476** | **6,233,202** | **2.2** |

Source: SciVal (June 18, 2025)

**Table S3. Shanghai Ranking of study and control group universities, 2018-2024**

| GROUP | 2018 | 2019 | 2020 | 2021 | 2022 | 2023 | 2024 |
|---|---|---|---|---|---|---|---|
| **STUDY** | | | | | | | |
| CU (India) | NR | NR | NR | NR | NR | NR | NR |
| Chitkara (India) | NR | NR | NR | NR | NR | NR | NR |
| GLA (India) | NR | NR | NR | NR | NR | NR | NR |
| KL (India) | NR | NR | NR | NR | NR | NR | NR |
| LPU (India) | NR | NR | NR | NR | NR | NR | NR |
| SIMATS (India) | NR | NR | NR | NR | NR | NR | 801-900 |
| UPES (India) | NR | NR | NR | NR | NR | NR | NR |
| LAU (Lebanon) | NR | NR | NR | NR | NR | NR | 501-600 |
| KKU (Saudi Arabia) | NR | NR | NR | 801-900 | 601-700 | 401-500 | 401-500 |
| KSU (Saudi Arabia) | 151-200 | 151-200 | 151-200 | 101-150 | 101-150 | 101-150 | 90 |
| NU (Saudi Arabia) | NR | NR | NR | NR | NR | NR | NR |
| PSAU (Saudi Arabia) | NR | NR | NR | NR | 801-900 | 501-600 | 601-700 |
| PNU (Saudi Arabia) | NR | NR | NR | NR | NR | 301-400 | 301-400 |
| TU (Saudi Arabia) | NR | NR | NR | 801-900 | 401-500 | 201-300 | 401-500 |
| UQU (Saudi Arabia) | NR | NR | NR | NR | NR | 801-900 | 701-800 |
| UOH (Saudi Arabia) | NR | NR | NR | NR | NR | NR | NR |
| UT (Saudi Arabia) | NR | NR | NR | NR | NR | NR | NR |
| UOS (United Arab Emirates) | NR | NR | NR | NR | NR | 901-1000 | 701-800 |
| **REGIONAL** | | | | | | | |
| IISc (India) | 401-500 | 401-500 | 501-600 | 401-500 | 301-400 | 301-400 | 401-500 |
| AUB (Lebanon) | 601-700 | 601-700 | 601-700 | 701-800 | 701-800 | 701-800 | 801-900 |
| KAUST (Saudi Arabia) | 201-300 | 201-300 | 201-300 | 201-300 | 201-300 | 201-300 | 201-300 |
| KU (United Arab Emirates) | NR | NR | NR | NR | 801-900 | 801-900 | 601-700 |
| **INTERNATIONAL** | | | | | | | |
| ETH Zurich | 19 | 19 | 20 | 21 | 20 | 20 | 21 |
| MIT | 4 | 4 | 4 | 4 | 3 | 3 | 3 |
| Princeton | 6 | 6 | 6 | 6 | 6 | 6 | 7 |
| UC Berkeley | 5 | 5 | 5 | 5 | 5 | 5 | 5 |

**Table S4. Changes in research output (count and top-100 world ranking) by subject category**

| Name | # of articles, 2018-2019 | # of articles, 2023-2024 | World ranking in # of articles, 2018-2019 | World ranking in # of articles, 2023-2024 |
|---|---|---|---|---|
| **Agricultural Sciences (n=1)** | | | | |
| **Study Group** | | | | |
| King Saud University | 327 | 1,316 | **82** | **9** |
| **International Group** | | | | |
| None ranked among the world's 100 most published | | | | |
| **Biology & Biochemistry (n=1)** | | | | |
| **Study Group** | | | | |
| King Saud University | 627 | 1,226 | **69** | **8** |
| **International Group** | | | | |
| Massachusetts Institute of Technology | 968 | 973 | **21** | **23** |
| University of California, Berkeley | 760 | 686 | **40** | **54** |
| ETH Zurich | 665 | 584 | **60** | **86** |
| **Chemistry (n=4)** | | | | |
| **Study Group** | | | | |
| King Saud University | 1,104 | 5,333 | **66** | **2** |
| King Khalid University | 313 | 1,993 | **552** | **33** |
| Saveetha Institute of Medical and Technical Sciences | 19 | 1,340 | **2001-3000** | **63** |
| Princess Nourah Bint Abdulrahman University | 114 | 1,286 | **1001-1500** | **68** |
| **International Group** | | | | |
| ETH Zurich | 1,366 | 1,118 | **43** | **84** |
| Massachusetts Institute of Technology | 1,263 | 939 | **51** | **Dropped out** |
| University of California, Berkeley | 1,140 | 909 | **63** | **Dropped out** |
| **Computer Science (n=4)** | | | | |
| **Study Group** | | | | |
| King Saud University | 375 | 749 | **57** | **45** |
| Lebanese American University | 37 | 613 | **1001-1500** | **59** |
| Princess Nourah Bint Abdulrahman University | 22 | 505 | **1501-2000** | **73** |
| Prince Sattam Bin Abdulaziz University | 32 | 473 | **1001-1500** | **87** |
| **International Group** | | | | |
| Massachusetts Institute of Technology | 472 | 482 | **41** | **82** |
| ETH Zurich | 362 | 447 | **62** | **91** |
| University of California, Berkeley | 291 | 244 | **99** | **Dropped out** |
| **Economics & Business (n=1)** | | | | |
| **Study Group** | | | | |
| Lebanese American University | 47 | 518 | **734** | **19** |
| **International Group** | | | | |
| Massachusetts Institute of Technology | 421 | 421 | **11** | **39** |
| University of California, Berkeley | 388 | 397 | **22** | **48** |
| ETH Zurich | 289 | 247 | **55** | **Dropped out** |
| **Engineering (n=3)** | | | | |
| **Study Group** | | | | |
| King Saud University | 794 | 3,453 | **145** | **31** |
| King Khalid University | 242 | 2,293 | **602** | **54** |
| Prince Sattam Bin Abdulaziz University | 148 | 1,667 | **910** | **82** |
| **International Group** | | | | |
| ETH Zurich | 1,248 | 1,465 | **75** | **Dropped out** |
| Massachusetts Institute of Technology | 1,604 | 1,245 | **43** | **Dropped out** |
| University of California, Berkeley | 1,132 | 857 | **86** | **Dropped out** |
| **Environment/Ecology (n=2)** | | | | |
| **Study Group** | | | | |
| King Saud University | 313 | 1,888 | **208** | **3** |
| Saveetha Institute of Medical and Technical Sciences | 7 | 659 | **3001-4000** | **70** |
| **International Group** | | | | |
| ETH Zurich | 914 | 836 | **19** | **40** |
| University of California, Berkeley | 865 | 776 | **21** | **49** |
| **Geosciences (n=0)** | | | | |
| **Study Group** | | | | |
| None ranked among the world's 100 most published | | | | |
| **International Group** | | | | |
| ETH Zurich | 1,353 | 1,433 | **10** | **12** |
| Massachusetts Institute of Technology | 628 | 567 | **53** | **75** |
| University of California, Berkeley | 655 | 504 | **49** | **97** |
| Princeton University | 527 | 501 | **82** | **99** |
| **Immunology (n=0)** | | | | |
| **Study Group** | | | | |
| None ranked among the world's 100 most published | | | | |
| **International Group** | | | | |
| Massachusetts Institute of Technology | 277 | 348 | **85** | **79** |
| **Materials Science (n=3)** | | | | |
| **Study Group** | | | | |
| King Saud University | 651 | 2,676 | **116** | **23** |
| King Khalid University | 288 | 1,345 | **344** | **65** |
| Saveetha Institute of Medical and Technical Sciences | 27 | 1,159 | **2001-3000** | **80** |
| **International Group** | | | | |
| Massachusetts Institute of Technology | 1,139 | 966 | **47** | **100** |
| ETH Zurich | 798 | 874 | **84** | **Dropped out** |
| University of California, Berkeley | 795 | 688 | **85** | **Dropped out** |
| **Mathematics (n=5)** | | | | |
| **Study Group** | | | | |
| King Saud University | 246 | 945 | **127** | **1** |
| Princess Nourah Bint Abdulrahman University | 19 | 607 | **1501-2000** | **12** |
| Prince Sattam Bin Abdulaziz University | 62 | 526 | **760** | **16** |
| King Khalid University | 75 | 489 | **654** | **22** |
| Lebanese American University | 4 | 294 | **2001-3000** | **98** |
| **International Group** | | | | |
| ETH Zurich | 427 | 458 | **20** | **29** |
| Massachusetts Institute of Technology | 529 | 435 | **8** | **33** |
| University of California, Berkeley | 452 | 427 | **18** | **34** |
| Princeton University | 404 | 392 | **28** | **45** |
| **Microbiology (n=1)** | | | | |
| **Study Group** | | | | |
| King Saud University | 107 | 305 | **190** | **33** |
| **International Group** | | | | |
| Massachusetts Institute of Technology | 267 | 235 | **30** | **61** |
| University of California, Berkeley | 216 | 202 | **58** | **85** |
| ETH Zurich | 180 | 141 | **82** | **Dropped out** |
| **Molecular Biology & Genetics (n=0)** | | | | |
| **Study Group** | | | | |
| None ranked among the world's 100 most published | | | | |
| **International Group** | | | | |
| Massachusetts Institute of Technology | 1,221 | 1,150 | **10** | **8** |
| University of California, Berkeley | 554 | 462 | **68** | **79** |
| ETH Zurich | 440 | 393 | **96** | **Dropped out** |
| **Neuroscience & Behavior (n=0)** | | | | |
| **Study Group** | | | | |
| None ranked among the world's 100 most published | | | | |
| **International Group** | | | | |
| Massachusetts Institute of Technology | 572 | 576 | **81** | **91** |
| **Pharmacology & Toxicology (n=5)** | | | | |
| **Study Group** | | | | |
| King Saud University | 510 | 1,346 | **23** | **1** |
| Prince Sattam Bin Abdulaziz University | 135 | 475 | **316** | **61** |
| King Khalid University | 101 | 439 | **444** | **69** |
| Saveetha Institute of Medical and Technical Sciences | 29 | 411 | **1001-1500** | **80** |
| Umm Al-Qura University | 88 | 348 | **510** | **100** |
| **International Group** | | | | |
| None ranked among the world's 100 most published | | | | |
| **Physics (n=3)** | | | | |
| **Study Group** | | | | |
| King Saud University | 326 | 1,202 | **286** | **33** |
| King Khalid University | 269 | 784 | **353** | **81** |
| Princess Nourah Bint Abdulrahman University | 21 | 678 | **2001-3000** | **97** |
| **International Group** | | | | |
| Massachusetts Institute of Technology | 1,921 | 1,705 | **9** | **15** |
| Princeton University | 1,388 | 1,234 | **22** | **29** |
| ETH Zurich | 1,218 | 1,128 | **26** | **39** |
| University of California, Berkeley | 1,209 | 1,054 | **29** | **46** |
| **Plant & Animal Science (n=1)** | | | | |
| **Study Group** | | | | |
| King Saud University | 426 | 1,712 | **128** | **5** |
| **International Group** | | | | |
| University of California, Berkeley | 545 | 428 | **78** | **Dropped out** |
| None ranked among the world's 100 most published | | | | |
| **Psychiatry/Psychology (n=0)** | | | | |
| **Study Group** | | | | |
| None ranked among the world's 100 most published | | | | |
| **International Group** | | | | |
| University of California, Berkeley | 489 | 434 | **88** | **Dropped out** |
| **Social Sciences, general (n=0)** | | | | |
| **Study Group** | | | | |
| None ranked among the world's 100 most published | | | | |
| **International Group** | | | | |
| University of California, Berkeley | 1,239 | 1,062 | **35** | **64** |
| **Space Science (n=0)** | | | | |
| **Study Group** | | | | |
| None ranked among the world's 100 most published | | | | |
| **International Group** | | | | |
| University of California, Berkeley | 1,130 | 1,254 | **6** | **9** |
| Massachusetts Institute of Technology | 761 | 1,083 | **19** | **14** |
| Princeton University | 783 | 1,015 | **17** | **15** |
| ETH Zurich | 351 | 362 | **71** | **81** |

**Data source:** Essential Science Indicators, via InCites (May 1, 2025). None of the regional control group universities appeared in the global top 100 most-published institutions by subject.

**Table S5. Illustrative cases of hyper-prolific authorship at study group universities and those listed in Table S1**

| Case Description | Case data |
|---|---|
| **Case 1:** A hyper-prolific author (HPA) affiliated with a study group university published 500 articles between 2021 and 2025, listing an average of 3.5 institutional affiliations per publication. Eighty percent of articles included 2 to 9 simultaneous affiliations, often spanning two to three countries. Nearly all the institutions involved exhibited bibliometric patterns consistent with the broader trends observed among the study group universities. | Middle East University Jordan 276<br>Universiti Sains Malaysia 246<br>Al Al-Bayt University 245<br>Applied Science Private University 243<br>Al-Ahliyya Amman University 241<br>Amman Arab University 178<br>Sunway University 109<br>Lebanese American University 107 |
| **Case 2:** This author began publishing three articles in 2019, sharply increasing to 37 publications in 2020. Between 2021 and 2024, the author averaged 123 articles per year. Sixty percent of these publications included 2 to 4 concurrent affiliations across 2-3 countries, with each affiliated institution appearing in over 100 articles. | 2024 152<br>2023 105<br>2022 93<br>2021 143<br>2020 37<br>2019 3 |
| **Case 3:** Initially averaging one publication per year over 14 years, this author's output increased dramatically after being appointed dean, reaching 230 articles in 2022 and 510 in 2023, with no co-authors from their home institution. In 2024, output fell to 54 publications, and no articles were published in 2025. The abrupt rise and subsequent decline raise questions about the sustainability and structure of such publication trajectories, particularly in leadership contexts. | 2024 55<br>2023 509<br>2022 229<br>2021 1<br>2014 1<br>2013 1<br>2010 3<br>2008 1 |

**Data Source:** Scopus (May 1, 2025).

**Table S6. Number and proportion of articles published in delisted journals, 2018-2019 and 2023-2024**

| | Articles in delisted journals | | % of total output | | World ranking in # of articles in delisted journals | |
|---|---|---|---|---|---|---|
| | 2018-2019 | 2023-2024 | 2018-2019 | 2023-2024 | 2018-2019 | 2023-2024 |
| **STUDY GROUP** | | | | | | |
| Chandigarh University | 299 | 293 | 57.4% | 6.5% | 4 | 25 |
| Chitkara University | 117 | 182 | 35.3% | 6.4% | 13 | 27 |
| GLA University | 99 | 180 | 25.8% | 5.9% | 23 | 36 |
| Koneru Lakshmaiah Education Foundation | 2550 | 576 | 77.3% | 15.1% | 2 | 2 |
| Lovely Professional University | 517 | 281 | 34.4% | 5.4% | 15 | 45 |
| Saveetha Institute of Medical and Technical Sciences | 2,421 | 874 | 79.7% | 8.4% | 1 | 14 |
| University of Petroleum and Energy Studies | 132 | 129 | 27.0% | 4.8% | 21 | 58 |
| Lebanese American University | 10 | 359 | 1.7% | 6.2% | 496 | 29 |
| King Khalid University | 236 | 657 | 11.4% | 5.9% | 44 | 34 |
| King Saud University | 697 | 986 | 8.2% | 3.6% | 64 | 79 |
| Najran University | 35 | 151 | 10.4% | 6.2% | 53 | 30 |
| Prince Sattam Bin Abdulaziz University | 163 | 683 | 13.1% | 8.0% | 36 | 16 |
| Princess Nourah Bint Abdulrahman University | 93 | 528 | 11.4% | 6.1% | 46 | 31 |
| Taif University | 121 | 385 | 12.1% | 7.5% | 40 | 20 |
| Umm Al-Qura University | 96 | 363 | 8.5% | 6.4% | 63 | 26 |
| University of Hail | 59 | 172 | 11.0% | 5.5% | 50 | 42 |
| University of Tabuk | 71 | 196 | 10.1% | 5.9% | 58 | 37 |
| University of Sharjah | 45 | 220 | 3.6% | 4.8% | 219 | 60 |
| **REGIONAL GROUP** | | | | | | |
| Indian Institute of Science, Bangalore | 37 | 39 | 0.9% | 0.9% | 710 | 633 |
| American University of Beirut | 38 | 27 | 1.7% | 1.3% | 503 | 438 |
| King Abdullah University of Science and Technology | 17 | 31 | 0.5% | 0.6% | 936 | 744 |
| Khalifa University of Science and Technology | 23 | 70 | 1.3% | 1.8% | 603 | 271 |
| **INTERNATIONAL GROUP** | | | | | | |
| Swiss Federal Institute of Technology Zurich | 30 | 23 | 0.2% | 0.2% | 997 | 995 |
| Massachusetts Institute of Technology | 46 | 28 | 0.3% | 0.2% | 991 | 991 |
| Princeton University | 10 | 9 | 0.1% | 0.1% | 1000 | 998 |
| University of California at Berkeley | 58 | 44 | 0.4% | 0.3% | 962 | 958 |

**Data Source:** Scopus and Web of Science (May 1, 2025).

**Table S7. Number, proportion, and world rank in retractions**

| | Articles retracted | | Retractions per 1,000 articles | | World ranking in retraction rate | |
|---|---|---|---|---|---|---|
| | 2018-19 | 2022-23 | 2018-19 | 2022-23 | 2018-19 | 2022-23 |
| **STUDY GROUP** | | | | | | |
| Chandigarh University | 0 | 38 | 0.0 | 11.1 | NA | 19 |
| Chitkara University | 0 | 30 | 0.0 | 17.7 | NA | 8 |
| GLA University | 0 | 38 | 0.0 | 16.4 | NA | 9 |
| Koneru Lakshmaiah Education Foundation | 5 | 52 | 1.5 | 19.2 | 162 | 4 |
| Lovely Professional University | 0 | 33 | 0.0 | 9.2 | NA | 32 |
| Saveetha Institute of Medical & Technical Sciences | 1 | 175 | 0.3 | 27.6 | 613 | 1 |
| University of Petroleum and Energy Studies | 4 | 40 | 8.2 | 14.2 | 26 | 35 |
| Lebanese American University | 0 | 16 | 0.0 | 5.0 | NA | 96 |
| King Khalid University | 1 | 108 | 0.5 | 10.8 | 464 | 21 |
| King Saud University | 14 | 222 | 1.6 | 10.9 | 145 | 20 |
| Najran University | 0 | 20 | 0.0 | 8.9 | NA | 34 |
| Prince Sattam Bin Abdulaziz University | 5 | 129 | 4.0 | 15.0 | 54 | 11 |
| Princess Nourah Bint Abdulrahman University | 2 | 79 | 2.4 | 9.9 | 90 | 27 |
| Taif University | 2 | 129 | 2.0 | 18.5 | 115 | 6 |
| Umm Al-Qura University | 1 | 51 | 0.9 | 8.6 | 257 | 36 |
| University of Hail | 0 | 28 | 0.0 | 10.0 | NA | 26 |
| University of Tabuk | 3 | 31 | 4.3 | 11.9 | 47 | 17 |
| University of Sharjah | 2 | 12 | 1.6 | 2.8 | 154 | 188 |
| **REGIONAL GROUP** | | | | | | |
| Indian Institute of Science, Bangalore | 2 | 3 | 0.5 | 0.6 | 463 | 443 |
| American University of Beirut | 0 | 1 | 0.0 | 0.4 | NA | 547 |
| King Abdullah University of Science and Technology | 0 | 5 | 0.0 | 1.1 | NA | 351 |
| Khalifa University of Science and Technology | 0 | 3 | 0.0 | 0.8 | NA | 392 |
| **INTERNATIONAL GROUP** | | | | | | |
| Swiss Federal Institute of Technology Zurich | 3 | 6 | 0.2 | 0.4 | 697 | 545 |
| Massachusetts Institute of Technology | 6 | 4 | 0.4 | 0.3 | 540 | 703 |
| Princeton University | 1 | 0 | 0.1 | 0.0 | 767 | NA |
| University of California at Berkeley | 8 | 5 | 0.6 | 0.3 | 407 | 602 |

**Data Source:** Medline, Retraction Watch, and Web of Science (June 1, 2025).

**Table S8. Proportion of articles ranked among the world's 2% most cited, 2018-2019 and 2023-2024**

| | # of articles ranked among world's top 2% most cited | | % of total research output ranked among the world's top 2% | | World ranking in % of top 2% articles | | Increase in output from 2018-19 to 2023-24 | |
|---|---|---|---|---|---|---|---|---|
| | **2018-19** | **2023-24** | **2018-19** | **2023-24** | **2018-19** | **2023-24** | **Top 2%** | **Overall** |
| **STUDY GROUP** | | | | | | | | |
| CU | 13 | 386 | 2.5% | 8.6% | 630 | 14 | 2869% | 760% |
| Chitkara | 12 | 204 | 3.6% | 7.1% | 370 | 35 | 1600% | 766% |
| GLA | 9 | 213 | 2.3% | 6.9% | 661 | 41 | 2267% | 702% |
| KL | 0 | 205 | 0.0% | 5.4% | 995 | 134 | 20500% | 512% |
| LPU | 53 | 349 | 3.5% | 6.8% | 390 | 47 | 558% | 243% |
| SIMATS | 64 | 872 | 2.1% | 8.4% | 712 | 15 | 1263% | 243% |
| UPES | 0 | 265 | 0.2% | 10.0% | 995 | 3 | 26500% | 444% |
| LAU | 14 | 783 | 2.4% | 13.5% | 646 | 1 | 5493% | 908% |
| KKU | 26 | 878 | 1.3% | 7.9% | 900 | 22 | 3277% | 436% |
| KSU | 336 | 1,141 | 4.0% | 4.2% | 277 | 280 | 240% | 220% |
| NU | 10 | 203 | 3.0% | 8.3% | 501 | 17 | 1930% | 545% |
| PSAU | 33 | 466 | 2.7% | 5.5% | 592 | 122 | 1312% | 584% |
| PNU | 15 | 462 | 1.8% | 5.3% | 781 | 140 | 2980% | 965% |
| TU | 14 | 295 | 1.4% | 5.7% | 872 | 105 | 2007% | 419% |
| UQU | 11 | 338 | 1.0% | 5.9% | 942 | 84 | 2973% | 407% |
| UOH | 15 | 170 | 2.8% | 5.5% | 565 | 127 | 1033% | 484% |
| UT | 26 | 182 | 3.7% | 5.4% | 351 | 130 | 600% | 375% |
| UOS | 53 | 271 | 4.2% | 5.9% | 232 | 87 | 411% | 267% |
| **Total/Average** | **706** | **7,683** | **2.5%** | **6.5%** | | | **988%** | **326%** |
| **REGIONAL GROUP** | | | | | | | | |
| IISc | 91 | 102 | 2.2% | 2.2% | 690 | 785 | 12% | 10% |
| AUB | 62 | 54 | 2.8% | 2.6% | 574 | 713 | -13% | -5% |
| KAUST | 361 | 419 | 10.0% | 8.7% | 2 | 12 | 16% | 34% |
| KU | 73 | 354 | 4.1% | 9.1% | 256 | 9 | 385% | 119% |
| **Total/Average** | **584** | **890** | **5.0%** | **6.0%** | | | **58%** | **31%** |
| **INT'L GROUP** | | | | | | | | |
| ETH Zurich | 841 | 724 | 6.7% | 5.3% | 26 | 144 | -14% | 9% |
| MIT | 1,685 | 1196 | 11.1% | 7.7% | 1 | 25 | -29% | 3% |
| Princeton | 529 | 395 | 7.7% | 5.3% | 10 | 139 | -25% | 8% |
| UC Berkeley | 1,112 | 798 | 7.8% | 5.7% | 9 | 101 | -28% | -2% |
| **Total/Average** | **3,928** | **2,722** | **8.5%** | **6.2%** | | | **-25%** | **4%** |

**Data Source:** SciVal (May 1, 2025).

**Table S9. Major citation contributors to the top 2% most cited articles (≥1% citation share)**

| | CU | Chitkara | GLA | KL | LPU | SIMATS | UPES | LAU | KKU | KSU | NU | PSAU | PNU | TU | UQU | UOH | UO |
|---|---|---|---|---|---|---|---|---|---|---|---|---|---|---|---|---|---|
| **CU** | 4% | Y | | | Y | Y | | Y | Y | Y | | | | | | | |
| **Chitkara** | Y | 16% | | | Y | Y | | | Y | Y | | | Y | | | | |
| **GLA** | Y | Y | 6% | | Y | Y | | | Y | Y | | Y | | Y | Y | | |
| **KL** | Y | Y | Y | 6% | Y | Y | | Y | Y | Y | | Y | Y | | | | |
| **LPU** | Y | Y | Y | | 7% | Y | | | Y | Y | | Y | | | | | |
| **SIMATS** | Y | Y | Y | | Y | 12% | | | Y | Y | | | | | | | |
| **UPES** | Y | Y | | | Y | Y | 3% | Y | Y | Y | | | | | | | |
| **LAU** | | Y | | | Y | Y | | 5% | Y | Y | | Y | Y | | | | |
| **KKU** | Y | Y | | | Y | Y | | Y | 8% | Y | | Y | Y | Y | Y | | Y |
| **KSU** | | Y | | | | Y | | | Y | 9% | | Y | Y | Y | | | |
| **NU** | | Y | | | | Y | | | Y | Y | 4% | Y | Y | Y | Y | | |
| **PSAU** | | Y | | | | Y | | Y | Y | Y | | 6% | Y | Y | Y | Y | |
| **PNU** | | Y | | | | Y | | Y | Y | Y | | Y | 7% | Y | Y | Y | Y |
| **TU** | | Y | | Y | Y | Y | | | Y | Y | | Y | Y | 5% | Y | Y | Y |
| **UQU** | | Y | | | Y | Y | | Y | Y | Y | | Y | Y | Y | 5% | Y | Y |
| **UOH** | | Y | | | Y | Y | | Y | Y | Y | | Y | Y | Y | Y | 5% | Y |
| **UOT** | | Y | | Y | | Y | | Y | Y | Y | | Y | Y | Y | Y | | 5% |
| **UOS** | | | | | | Y | | Y | Y | Y | | Y | | | | | |
| | **IISc** | **AUB** | **KAUST** | **KU** | | | | | | | | | | | | | |
| **IISc** | 3% | | | | | | | | | | | | | | | | |
| **AUB** | | 1% | | | | | | | | | | | | | | | |
| **KAUST** | | | 4% | | | | | | | | | | | | | | |
| **KU** | | | | 4% | | | | | | | | | | | | | |
| | **ETH** | **MIT** | **Princeton** | **UCB** | | | | | | | | | | | | | |
| **ETH** | 4% | | Y | | | | | | | | | | | | | | |
| **MIT** | | 3% | Y | Y | | | | | | | | | | | | | |
| **Princeton** | | | 4% | Y | | | | | | | | | | | | | |
| **UCB** | | Y | Y | 4% | | | | | | | | | | | | | |

**This table should be read as follows: the main citation contributors to CU (Chandigarh University's) top 2% most cited articles are CU itself (with a self-citation rate of 4%), six fellow study group institutions, and one other university beyond the study group. Data Source:** SciVal (May 1, 2025).

**Table S10. Number of institutions appearing as major co-authoring partners (≥2% co-authorship rate)**

| | Count of major collaborating institutions, 2018-2019 | Count of major collaborating institutions, 2023-2024 | New or intensified collaborators* |
|---|---|---|---|
| **STUDY GROUP** | | | |
| Chandigarh University | 5 | 20 | 17 |
| Chitkara University | 9 | 26 | 22 |
| GLA University | 6 | 23 | 22 |
| Koneru Lakshmaiah Education Foundation | 2 | 20 | 15 |
| Lovely Professional University | 4 | 19 | 13 |
| Saveetha Institute of Medical and Technical Sciences | 1 | 22 | 20 |
| University of Petroleum and Energy Studies | 5 | 23 | 18 |
| Lebanese American University | 8 | 42 | 39 |
| King Khalid University | 12 | 27 | 11 |
| King Saud University | 7 | 7 | 3 |
| Najran University | 13 | 26 | 14 |
| Prince Sattam Bin Abdulaziz University | 13 | 33 | 12 |
| Princess Nourah Bint Abdulrahman University | 16 | 40 | 16 |
| Taif University | 22 | 29 | 8 |
| Umm Al-Qura University | 14 | 31 | 12 |
| University of Hail | 20 | 28 | 6 |
| University of Tabuk | 18 | 34 | 18 |
| University of Sharjah | 9 | 19 | 8 |
| **Average per institution** | 10 | 26 | 15 |
| **REGIONAL GROUP** | | | |
| Indian Institute of Science, Bangalore | 3 | 2 | 0 |
| American University of Beirut | 4 | 12 | 1 |
| King Abdullah University of Science and Technology | 1 | 0 | 0 |
| Khalifa University of Science and Technology | 4 | 9 | 2 |
| **Average per institution** | 3 | 6 | 1 |
| **INTERNATIONAL GROUP** | | | |
| Swiss Federal Institute of Technology Zurich | 8 | 8 | 0 |
| Massachusetts Institute of Technology | 8 | 20 | 0 |
| Princeton University | 17 | 29 | 0 |
| University of California at Berkeley | 12 | 15 | 0 |
| **Average per institution** | 12 | 18 | 0 |

***Institutions that are new or have increased their collaboration activity over five times compared to 2018-2019 levels.**
Data Source: SciVal (May 1, 2025).

**Table S11. Major collaborators of the study group universities (≥2% collaboration rate), 2018-2019 vs. 2023-2024**

| | CU | Chitkara | GLA | KL | LPU | SIMATS | UPES | LAU | KKU | KSU | NU | PSAU | PNU | TU | UQU | UOH | UOT | UOS |
|---|---|---|---|---|---|---|---|---|---|---|---|---|---|---|---|---|---|---|
| **2018-2019** | | | | | | | | | | | | | | | | | | |
| **CU** | | | | | | | | | | | | | | | | | | |
| **Chitkara** | | | | | | | | | | | | | | | | | | |
| **GLA** | | | | | | | | | | | | | | | | | | |
| **KL** | | | | | | | | | | | | | | | | | | |
| **LPU** | | | | | | | | | | | | | | | | | | |
| **SIMATS** | | | | | | | | | | | | | | | | | | |
| **UPES** | | | | | Y | | | | | | | | | | | | | |
| **LAU** | | | | | | | | | | | | | | | | | | |
| **KKU** | | | | | | | | | | Y | | | | | | | | |
| **KSU** | | | | | | | | | | | | **Y** | **Y** | | | | | |
| **NU** | | | | | | | | | Y | Y | | | | | | | | |
| **PSAU** | | | | | | | | | | **Y** | | | | | | | | |
| **PNU** | | | | | | | | | Y | **Y** | | | | **Y** | Y | | | |
| **TU** | | | | | | | | | | Y | | | **Y** | | **Y** | | | |
| **UQU** | | | | | | | | | | Y | | | | **Y** | | | | |
| **UOH** | | | | | | | | | Y | Y | | | | | Y | | | |
| **UOT** | | | | | | | | | Y | Y | | | | | | | | |
| **UOS** | | | | | | | | | | Y | | | | | | | | |
| **2023-2024** | | | | | | | | | | | | | | | | | | |
| **CU** | | **Y** | **Y** | | **Y** | **Y** | **Y** | **Y** | Y | Y | | Y | | | | | | |
| **Chitkara** | **Y** | | **Y** | | **Y** | **Y** | **Y** | **Y** | Y | Y | | | | | | | | |
| **GLA** | **Y** | **Y** | | Y | **Y** | **Y** | | | Y | Y | | **Y** | | | | | | |
| **KL** | Y | | | | | **Y** | | | | Y | | | | | | | | |
| **LPU** | **Y** | **Y** | **Y** | | | **Y** | | **Y** | Y | Y | | | | | | | | |
| **SIMATS** | **Y** | **Y** | **Y** | **Y** | **Y** | | | **Y** | **Y** | Y | | **Y** | | | | | | |
| **UPES** | **Y** | **Y** | | | | Y | | **Y** | | Y | | Y | | | | | | |
| **LAU** | **Y** | **Y** | | | **Y** | **Y** | **Y** | | **Y** | Y | | Y | Y | | **Y** | | | **Y** |
| **KKU** | | | | | | **Y** | | **Y** | | | | **Y** | **Y** | **Y** | **Y** | | | |
| **KSU** | | | | | | | | | | | | | | | | | | |
| **NU** | | | | | | | | | | | | **Y** | Y | **Y** | **Y** | **Y** | Y | |
| **PSAU** | | | **Y** | | | **Y** | | | **Y** | | **Y** | | **Y** | **Y** | **Y** | | | |
| **PNU** | | | | | | | | | **Y** | | | **Y** | | | | **Y** | **Y** | |
| **TU** | | | | | | Y | | | **Y** | | **Y** | **Y** | | | | | **Y** | |
| **UQU** | | | | | | Y | | **Y** | **Y** | | **Y** | **Y** | Y | | | | **Y** | |
| **UOH** | | | | | | | | | | | **Y** | | **Y** | | | | **Y** | |
| **UOT** | | | | | | | | Y | Y | | | | **Y** | **Y** | **Y** | **Y** | | |
| **UOS** | | | | | | | | **Y** | | | | | Y | | | | | |

**Institutions marked in red, bolded font are coauthors for** at least 2% of each other's publications**. Data Source:** SciVal (May 1, 2025).

**Table S12. Research Integrity Risk Index ($RI^2$) global ranking: red-flagged and study and control group universities**

| Scopus Name | Country | Group | Rank by share in delisted journals | Rank by retraction rate | $RI^2$ Total Score | $RI^2$ World Rank | Shanghai Ranking 2024 |
|---|---|---|---|---|---|---|---|
| Koneru Lakshmaiah Education Foundation | India | Study | 2 | 4 | 0.838 | 1 | |
| Jawaharlal Nehru Technological University Hyderabad | India | | 3 | 3 | 0.807 | 2 | |
| Saveetha Institute of Medical and Technical Sciences | India | Study | 14 | 1 | 0.772 | 3 | 801-900 |
| Anna University | India | | 9 | 2 | 0.761 | 4 | |
| Dr. A.P.J. Abdul Kalam Technical University | India | | 6 | 5 | 0.702 | 5 | |
| University of Pune | India | | 1 | 30 | 0.675 | 6 | |
| Visvesvaraya Technological University | India | | 5 | 14 | 0.590 | 7 | |
| Taif University | Saudi Arabia | Study | 20 | 6 | 0.576 | 8 | 401-500 |
| SRM Institute of Science and Technology | India | | 13 | 10 | 0.551 | 9 | 701-800 |
| Prince Sattam Bin Abdulaziz University | Saudi Arabia | Study | 17 | 11 | 0.531 | 10 | 601-700 |
| Chitkara University | India | Study | 27 | 8 | 0.528 | 11 | |
| GLA University | India | Study | 37 | 9 | 0.486 | 12 | |
| Gandhi Institute of Technology and Management | India | | 7 | 51 | 0.483 | 13 | |
| Al Jouf University | Saudi Arabia | | 16 | 22 | 0.453 | 14 | |
| The University of Lahore | Pakistan | | 47 | 12 | 0.445 | 15 | 901-1000 |
| University of Jeddah | Saudi Arabia | | 18 | 24 | 0.441 | 16 | |
| Symbiosis International University | India | | 8 | 72 | 0.438 | 17 | |
| Hainan Medical University | China | | 88 | 7 | 0.430 | 18 | |
| Government College University Faisalabad | Pakistan | | 44 | 13 | 0.421 | 19 | |
| Chandigarh University | India | Study | 25 | 19 | 0.414 | 20 | |
| University of Petroleum and Energy Studies | India | Study | 56 | 13 | 0.416 | 21 | |
| Universitas Airlangga | Indonesia | | 4 | 248 | 0.413 | 22 | |
| University of Agriculture Faisalabad | Pakistan | | 36 | 16 | 0.405 | 23 | |
| University of Tabuk | Saudi Arabia | Study | 35 | 17 | 0.404 | 24 | |
| King Faisal University | Saudi Arabia | | 15 | 56 | 0.390 | 25 | 801-900 |
| King Khalid University | Saudi Arabia | Study | 34 | 21 | 0.387 | 26 | 401-500 |
| Vellore Institute of Technology | India | | 52 | 15 | 0.382 | 27 | 501-600 |
| Princess Nourah Bint Abdulrahman University | Saudi Arabia | Study | 32 | 27 | 0.376 | 28 | 301-400 |
| Jazan University | Saudi Arabia | | 23 | 45 | 0.366 | 29 | 901-1000 |
| Umm Al-Qura University | Saudi Arabia | Study | 28 | 36 | 0.362 | 30 | 701-800 |
| University of the Punjab | Pakistan | | 22 | 50 | 0.361 | 31 | 701-800 |
| Najran University | Saudi Arabia | Study | 31 | 34 | 0.360 | 32 | |
| University of Hail | Saudi Arabia | Study | 42 | 26 | 0.359 | 33 | |
| Qassim University | Saudi Arabia | | 21 | 62 | 0.354 | 34 | 701-800 |
| Universitas Hasanuddin | Indonesia | | 10 | 369 | 0.348 | 35 | |
| Universiti Teknologi MARA | Malaysia | | 11 | 273 | 0.347 | 36 | |
| Lovely Professional University | India | Study | 45 | 32 | 0.342 | 37 | |
| Menoufia University | Egypt | | 33 | 49 | 0.330 | 38 | |
| Islamia University | Pakistan | | 68 | 23 | 0.328 | 39 | |
| University of Baghdad | Iraq | | 12 | 353 | 0.324 | 40 | |
| King Saud University | Saudi Arabia | Study | 79 | 20 | 0.313 | 41 | 90 |
| Universiti Teknologi Malaysia | Malaysia | | 19 | 140 | 0.311 | 42 | |
| COMSATS University Islamabad | Pakistan | | 48 | 48 | 0.309 | 43 | 601-700 |
| Taibah University | Saudi Arabia | | 24 | 109 | 0.295 | 44 | |
| Al-Imam Muhammad Ibn Saud Islamic University | Saudi Arabia | | 49 | 64 | 0.292 | 45 | |
| Hebei Medical University | China | | 83 | 28 | 0.292 | 46 | 801-900 |
| Lebanese American University | Lebanon | Study | 31 | 96 | 0.291 | 47 | 501-600 |
| King Abdulaziz University | Saudi Arabia | | 57 | 54 | 0.282 | 48 | 201-300 |
| Xuzhou Medical University | China | | 165 | 18 | 0.281 | 49 | 801-900 |
| University of Technology- Iraq | Iraq | | 53 | 85 | 0.262 | 50 | |
| University of Sharjah | United Arab Emirates | Study | 58 | 188 | 0.204 | 77 | 701-800 |
| Khalifa University of Science and Technology | United Arab Emirates | Regional | 272 | 392 | 0.072 | 365 | 601-700 |
| American University of Beirut | Lebanon | Regional | 420 | 535 | 0.049 | 500 | 801-900 |
| King Abdullah University of Science and Technology | Saudi Arabia | Regional | 741 | 351 | 0.039 | 579 | 201-300 |
| Indian Institute of Science Bangalore | India | Regional | 631 | 443 | 0.038 | 585 | 401-500 |
| University of California at Berkeley | United States | International | 953 | 602 | 0.015 | 912 | 5 |
| Swiss Federal Institute of Technology Zurich | Switzerland | International | 994 | 545 | 0.012 | 955 | 21 |
| Massachusetts Institute of Technology | United States | International | 991 | 703 | 0.009 | 982 | 3 |
| Princeton University | United States | International | 996 | 926 | 0.002 | 999 | 7 |

Each institution's $RI^2$ score reflects the sum of its rank in proportion of publications in journals delisted by Scopus or Web of Science (2023-2024), and (2) retraction rate per 1,000 publications (2022-2023). Lower scores indicate higher bibliometric risk. Institutions are ranked from highest to lowest risk (Rank 1 = highest risk). Data were sourced from Retraction Watch (as of May 13, 2025). Only universities among the world's top 1,000 most published in 2023-2024 are included. The complete list is available at https://sites.aub.edu.lb/lmeho/ri2/.

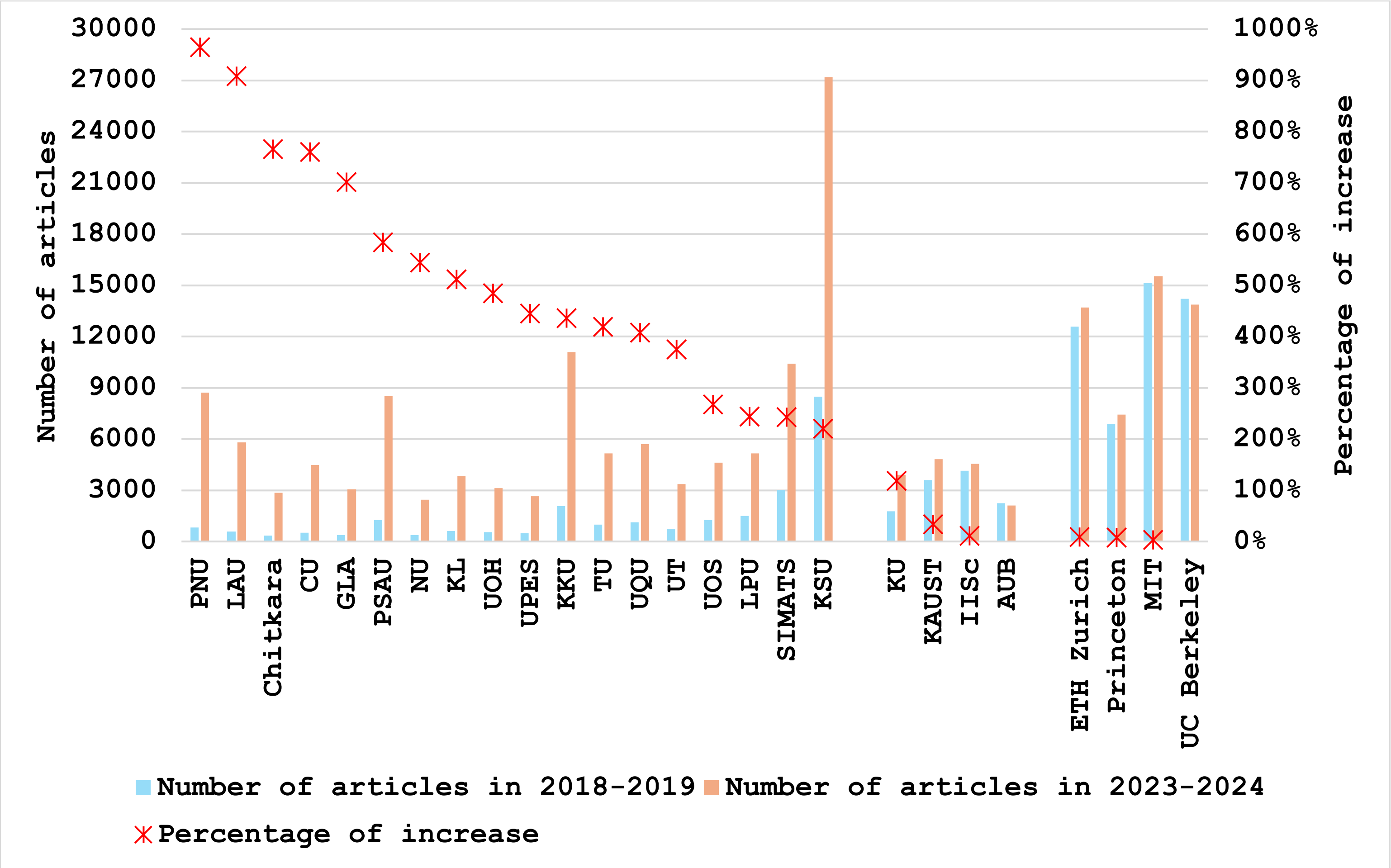

**Fig. S1. Extent of research growth among universities in the study, regional, and international groups between 2018-2019 and 2023-2024.** A blank space separates the study group (left) from the regional group (middle) and international group (right). The increase percentages for AUB and UC Berkeley do not appear because they dropped in research output by 6% and 3%, respectively. **Data Source:** SciVal (May 1, 2025). The data for KL and NU in this Figure are based on InCites.

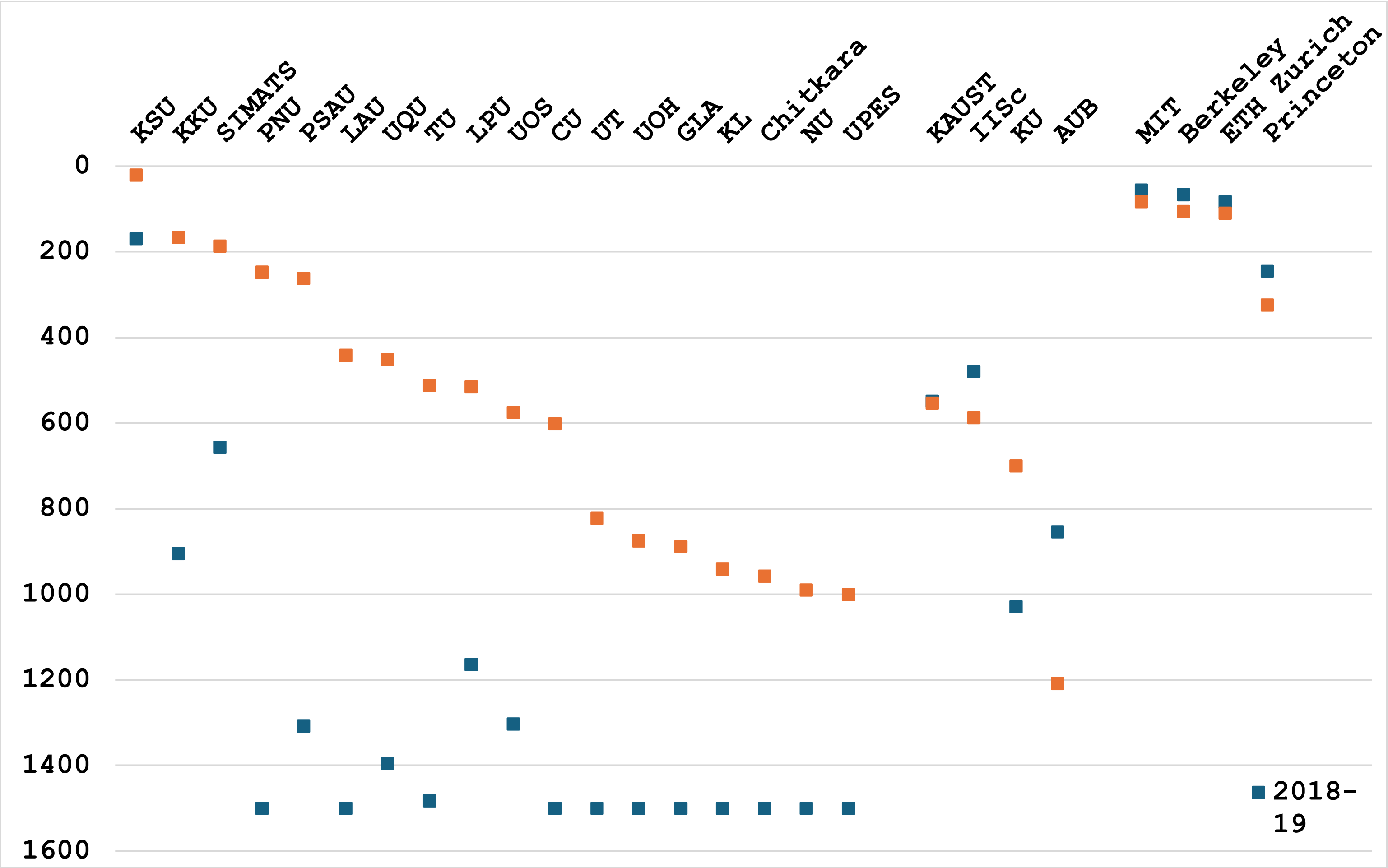

**Fig. S2. Changes in global research output rankings between 2018-2019 and 2023-2024 for universities in the study, regional, and international groups. Ten u**niversities ranked below 1,500 are depicted at 1,500th for visualization purposes**. Data Source:** SciVal (May 1, 2025). The data for KL and NU in this figure are based on InCites.

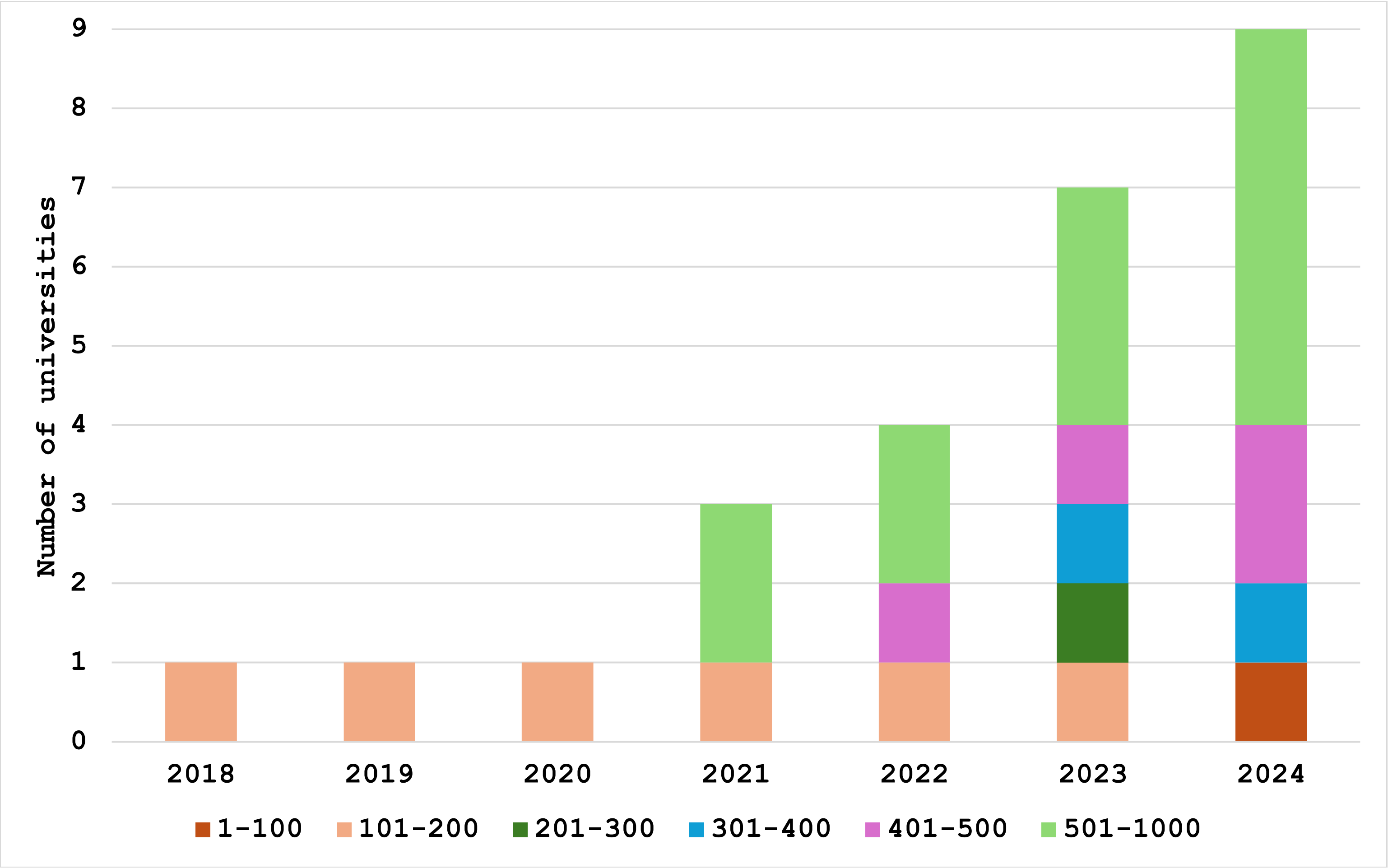

**Fig. S3. Study Group Institutions by Shanghai Ranking Tier (2018-2024)**

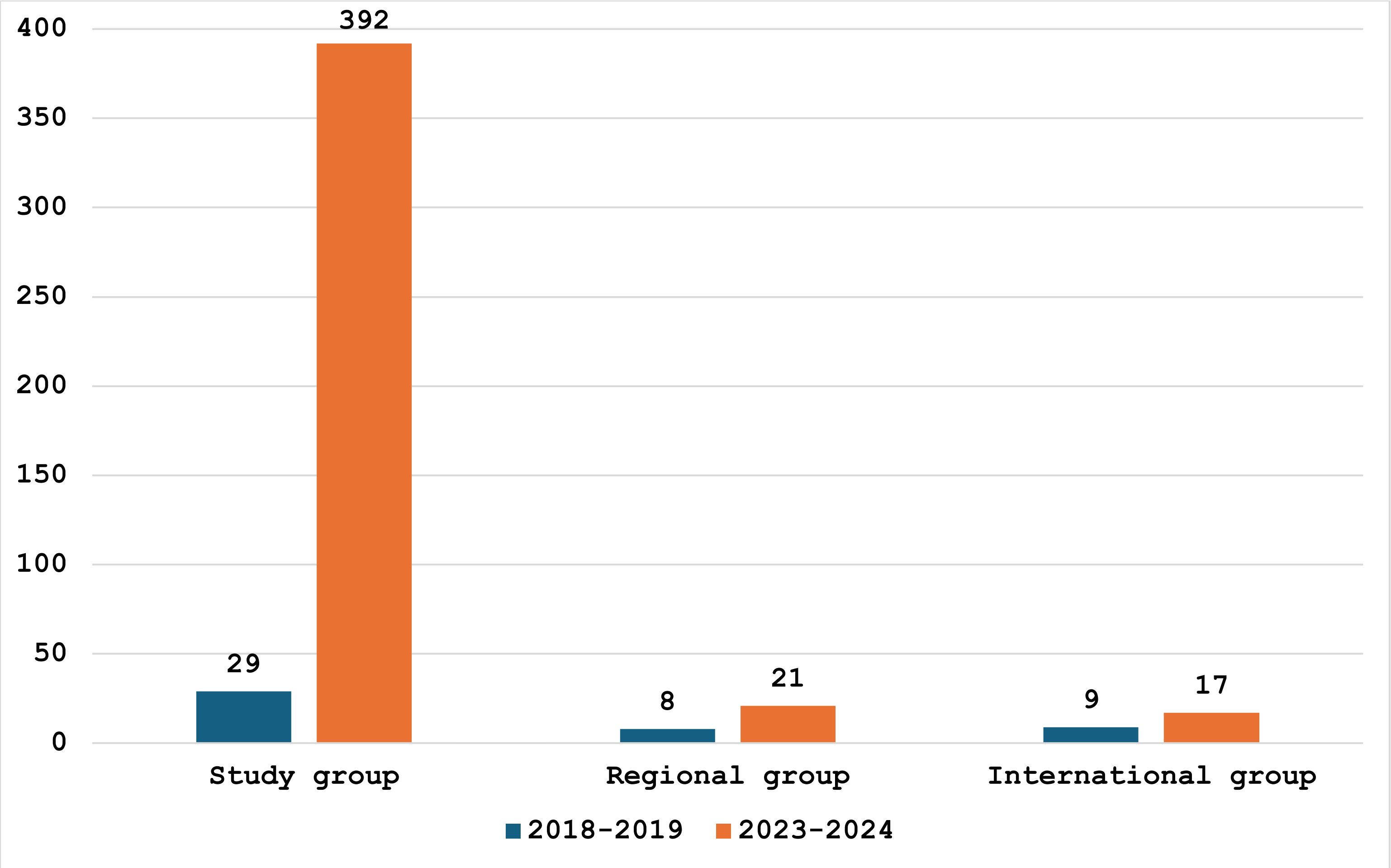

**Fig. S4. Increase in hyper-prolific authors among study and control group universities, 2018-2019 vs. 2023-2024. Data Source: SciVal (May 1, 2025).**